\documentclass{article}

\PassOptionsToPackage{numbers, compress}{natbib}
\usepackage[main, final]{neurips_2026}

\usepackage[utf8]{inputenc} %
\usepackage[T1]{fontenc}    %
\usepackage{hyperref}       %
\usepackage{url}            %
\usepackage{booktabs}       %
\usepackage{amsfonts}       %
\usepackage{nicefrac}       %
\usepackage{microtype}      %
\usepackage{xcolor}         %
\usepackage{multirow}
\usepackage{graphicx}
\usepackage{wrapfig}
\usepackage{amsmath}
\usepackage{amssymb}
\usepackage{amsthm}
\usepackage{colortbl}
\usepackage{pifont}
\usepackage{enumitem}
\usepackage{algorithm}
\usepackage{algpseudocode}

\definecolor{lightcyan}{rgb}{0.90, 0.97, 1.0}

\newcommand{\ie}{\textit{i}.\textit{e}., }

\newcommand{\ms}[2]{%
  \begin{tabular}[c]{@{}c@{}}
    #1\\[-2pt] %
    {\tiny $\pm$#2} %
  \end{tabular}%
}

\title{MUTE: Multi-Level Alignment Uncoupling Against Talking-Head Exploitation for Voice Protection}

\author{%
  \textbf{Donghyun Kim}$^{*,1}$,
  \textbf{Jin Hong}$^{*,2}$,
  \textbf{Seungmin Kim}$^{*,1,3}$, \\
  \textbf{Dain Kim}$^{1}$,
  \textbf{Junseok Kwon}$^{\dagger,2}$,
  \textbf{Daeseon Choi}$^{\dagger,1}$ \\
  $^{1}$Soongsil University \quad $^{2}$Chung-Ang University \quad $^{3}$AIM Intelligence \\
  \texttt{\{kmit16, kimsm0802, dikim420\}@soongsil.ac.kr} \quad \\
  \texttt{\{jindl465, jskwon\}@cau.ac.kr} \quad \texttt{sunchoi@ssu.ac.kr} \\
  \small{$^*$Equal contribution \quad $^{\dagger}$Corresponding authors}
}

\begin{document}

\maketitle

\begin{abstract}

Talking-head generation models, which synthesize realistic facial animations from audio, are increasingly vulnerable to misuse in multimodal deepfake scenarios. Protecting such systems remains challenging, as talking-head models fundamentally rely on precise audio–visual alignment, while effective audio protection methods remain largely underexplored. In this work, we propose MUTE, an audio protection framework that uncouples the underlying audio–visual alignment through a multi-level strategy. Specifically, MUTE combines (1) \emph{Alignment-level Disruption}, which directly perturbs cross-modal attention to disrupt structural alignment, and (2) \emph{Representation-level Degradation}, which perturbs temporal audio embeddings to degrade their representations. To improve robustness, perturbations are constrained in the STFT domain to high-energy regions, making them resistant to post-processing, denoising, and adaptive attacks. An optional speaker-level objective further mitigates potential bypass via TTS-based resynthesis. Extensive experiments demonstrate that MUTE consistently degrades lip synchronization across both white-box and black-box talking-head models while preserving perceptual audio quality. The proposed method remains effective under real-world transformations and can be combined with image-based approaches to provide stronger multimodal protection. The source code and project page are available at \url{https://github.com/jindl465/MUTE}.

\end{abstract}
\section{Introduction}
\label{sec:intro}

Rapid advances in generative models~\cite{sadtalker, adnerf, faceformer, audio2head, hallo, hallo2, hallo3, genefacepp, ditto, wav2lip, makeittalk} have significantly improved the realism of talking-head generation systems that jointly leverage audio and visual information. As generative AI continues to evolve, deepfake technologies have progressed from unimodal settings, such as image or speech manipulation, to more sophisticated multimodal scenarios that combine both modalities. This evolution has amplified the potential for misuse, raising growing concerns over multimodal deepfake attacks and motivating the need for data protection methods that go beyond conventional unimodal settings. However, unlike unimodal generation, talking-head models critically rely on precise audio-visual alignment between speech and facial motion. Existing protection methods designed for image~\cite{faceshield, nullswap, dualdefense} or speech deepfakes~\cite{antifake, roco, e2evguard} largely overlook such cross-modal coupling and therefore fail to target the alignment mechanism on which talking-head generation depends.

Recently, data protection methods specifically targeting talking-head generation models have been proposed. In particular, an approach that introduces perturbations to images~\cite{silencer} has demonstrated the potential to degrade generation quality from the visual modality. Furthermore, a method that simultaneously applies perturbations to both image and audio inputs~\cite{syncbreaker} has shown that stronger suppression effects can be achieved by directly disrupting multimodal alignment. These findings suggest that effective data protection for talking-head generation models should be approached from a multimodal perspective rather than being limited to unimodal settings.

\begin{figure*}[t]
    \begin{minipage}[t]{\linewidth}
	\centering
  \includegraphics[width=1.0\linewidth]{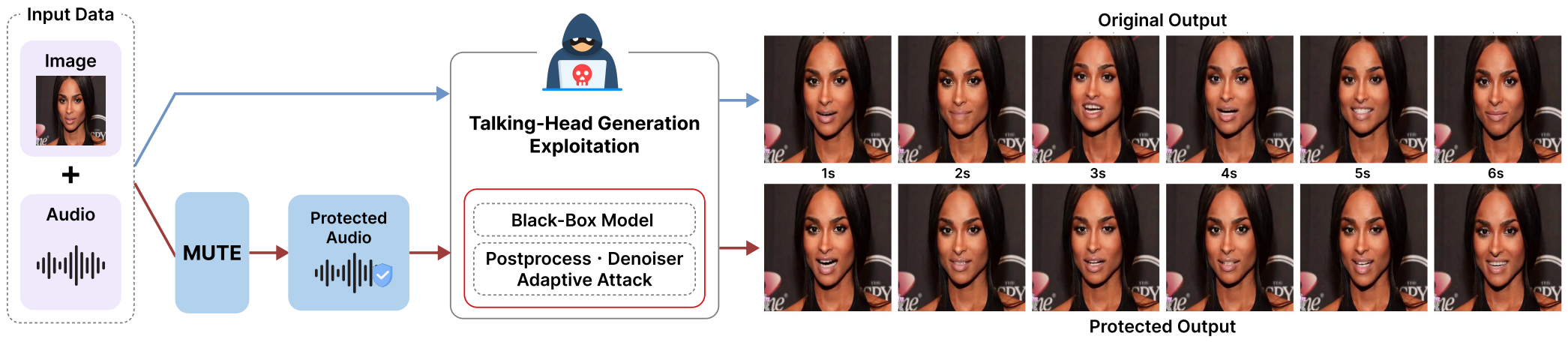}
    \end{minipage}
    \vspace{-5mm}
  \caption{\textbf{Overview of MUTE.} This conceptual overview shows how MUTE preemptively protects audio data from malicious synthesis. Even in real-world scenarios, the resulting outputs remain visually unsynchronized and unusable due to severe audio-visual misalignment.}
  \label{fig:overview}
  \vspace{-3mm}
\end{figure*}

Despite these advances, existing approaches still exhibit several important limitations. In realistic security scenarios, protecting against the misuse of talking-head generation must satisfy multiple requirements. First, since the specific talking-head generation model used by an attacker is unknown in advance, strong transferability under black-box settings is essential. Second, robustness is critical, as attackers can apply post-processing techniques, denoising methods, or adaptive attack to remove perturbations and neutralize their effects. 
In particular, for audio, an attacker can neutralize protection by extracting the speaker identity from protected speech and resynthesizing it via a text-to-speech (TTS) model. Despite these challenges, existing methods do not address these challenges systematically. Moreover, modality-specific considerations, especially for audio, remain relatively underexplored.

To address these limitations, we propose \textbf{MUTE} (\textbf{M}ulti-level alignment \textbf{U}ncoupling against \textbf{T}alking-head \textbf{E}xploitation), an audio protection framework designed to achieve strong transferability under black-box settings while being robust to diverse real-world transformations. The key idea of MUTE is to uncouple audio-visual alignment through a multi-level strategy operating at both alignment structure and representation levels.

First, to achieve \textbf{Alignment-level Disruption}, we employ a \textbf{Target Attention KL Loss} that perturbs the shared audio-visual alignment structure by guiding the attention map toward a target distribution. Subsequently, to achieve \textbf{Representation-level Degradation}, we introduce a \textbf{Frame-wise Embedding Cosine Loss}, which decomposes temporal audio embeddings and effectively degrades the embedding extraction process. Together, these objectives degrade both the structural mechanism and the input representations responsible for alignment. Furthermore, to enhance robustness, we constrain perturbations in the STFT domain to concentrate in high-energy regions, making them less susceptible to removal by post-processing, denoising and adaptive attack. In addition, we optionally incorporate a speaker-level perturbation objective to reduce identity consistency, thereby mitigating potential bypass via TTS-based resynthesis.

Fig.\ref{fig:overview} illustrates the effectiveness of MUTE under these scenarios. Through this design, MUTE consistently degrades lip synchronization across black-box models and remains effective under real-world transformation scenarios, including post-processing, denoising, adaptive attack and TTS-based resynthesis, while preserving perceptual audio quality. Moreover, MUTE operates solely on audio inputs and can be combined with image-based approaches to enable stronger multimodal protection.

Our main contributions are summarized as follows:
\begin{itemize}[leftmargin=*]
    \item We propose {MUTE}, a multi-level audio protection framework that combines \emph{Alignment-level Disruption} and \emph{Representation-level Degradation}, effectively uncoupling audio-visual alignment and mitigating the misuse of talking-head generation systems.

    \item We show that {MUTE} is complementary to existing image-based methods, achieving stronger protection when both modalities are jointly considered.
    
    \item We demonstrate that {MUTE} achieves strong performance under black-box settings by targeting model-agnostic audio-visual alignment.

    \item We introduce a robustness-aware perturbation strategy in the STFT domain that concentrates  perturbations to high-energy regions, improving resistance to post-processing, denoising, adaptive attack and extend {MUTE} with a speaker-level objective to mitigate TTS-based resynthesis attacks.

\end{itemize}

\section{Related Work}
\label{sec:related}

\noindent\textbf{Audio Adversarial Attacks.}
Audio adversarial attacks against Automatic Speech Recognition (ASR) systems have progressed from waveform-level optimization \cite{carlini}, to perceptually-aware perturbations leveraging psychoacoustic masking~\cite{psychoacoustic, qin}, and more recently to transferable attacks designed for black-box settings~\cite{aro, sago}. These methods have been shown to affect self-supervised models like Wav2Vec 2.0~\cite{wav2vec} and Whisper~\cite{whisper}, degrading learned representations and downstream performance~\cite{sok, robust}. Since talking-head generation relies on the linguistic and phonetic content of speech to drive facial motion, ASR-based adversarial attacks provide a natural starting point for evaluating audio-domain perturbations in this setting. We therefore include \emph{C\&W}~\cite{carlini}, \emph{Psychoacoustic}~\cite{psychoacoustic}, \emph{Qin et al.}~\cite{qin}, \emph{ARO}~\cite{aro}, and \emph{SAGO}~\cite{sago} as baselines in our experiments.

\noindent\textbf{Talking-Head Generation.}
Talking-head generation aims to synthesize realistic facial animations driven by control signals, most commonly audio. Early approaches were primarily based on GANs~\cite{wav2lip, makeittalk}, focusing on accurate lip synchronization and identity preservation. Subsequent methods introduced structured motion representations, such as 3D priors~\cite{sadtalker} and disentangled facial dynamics~\cite{faceformer, audio2head}, as well as neural radiance fields~\cite{adnerf}, to improve motion diversity and controllability. More recently, diffusion-based models~\cite{hallo, hallo2, hallo3, genefacepp, ditto} have emerged as the dominant paradigm, achieving superior visual quality and temporal coherence. Across these paradigms, audio remains the primary driving modality, enabling highly realistic and synchronized facial animations. However, such strong dependency on audio also introduces potential vulnerabilities, as manipulated or adversarial inputs can significantly affect the generation process. This observation has motivated a growing body of work on adversarial attacks and defenses for talking-head generation.

\noindent\textbf{Adversarial Attacks on Talking-Head Models.}
Adversarial attack methods on talking-head generation are still nascent, with only a few works exploring this direction. Silencer~\cite{silencer} perturbs the input portrait but is often less effective in audio-driven settings where facial motion is primarily controlled by speech. To address this limitation, audio-level attacks introduce perturbations to the speech signal to disrupt audio-visual synchronization. More recently, SyncBreaker~\cite{syncbreaker} enhances attack effectiveness by jointly leveraging audio perturbations with visual perturbations, enabling stronger disruption across different talking-head models. We also include \emph{CAF}~\cite{syncbreaker}, the audio protection method of SyncBreaker, as a baseline. Despite these advances, existing approaches still struggle to achieve both transferability under black-box settings and robustness against post-processing, denoising, adaptive attacks or TTS-based bypass — limitations that motivate our multi-level audio-based protection framework.
\section{Problem Formulation}
\label{sec:Problem}

We formulate audio protection against talking-head generation as a proactive defense problem, where adversarial perturbations are intentionally introduced to prevent unauthorized audio-driven synthesis. Given a clean audio signal $x$, we aim to generate an adversarial audio $x^{adv} = x + \delta$, where $\delta$ denotes a small perturbation. The goal is to degrade the synchronization quality of the generated talking-head output while preserving perceptual similarity to the original audio. In this setting, only the audio modality is modified, while the visual input remains unchanged.
To ensure perceptual imperceptibility, the perturbation is constrained by $\|x^{adv} - x\|_\infty \le \epsilon$.

The primary objective is to disrupt lip synchronization between audio and visual modalities. In addition, we consider secondary effects such as degradation in per-frame visual quality, temporal coherence, and facial expression consistency. We characterize the attack as a disruption of audio-visual alignment. Specifically, the perturbations first disrupt the cross-modal attention structure that governs audio-visual alignment and then degrade temporal audio representations. The detailed design is described in Section~\ref{sec:method}.

We assume a transfer-based black-box setting, where adversarial perturbations are generated without access to the architecture or parameters of the target model. Instead, the adversarial audio is generated on a surrogate model and directly evaluated on unseen talking-head models. Furthermore, the adversarial audio may undergo post-processing, denoising, adaptive attack and TTS-based resynthesis, necessitating robustness against such operations.
\section{Method}
\label{sec:method}

\begin{figure*}[t]
    \begin{minipage}[t]{\linewidth}
	\centering
  \includegraphics[width=1.0\linewidth]{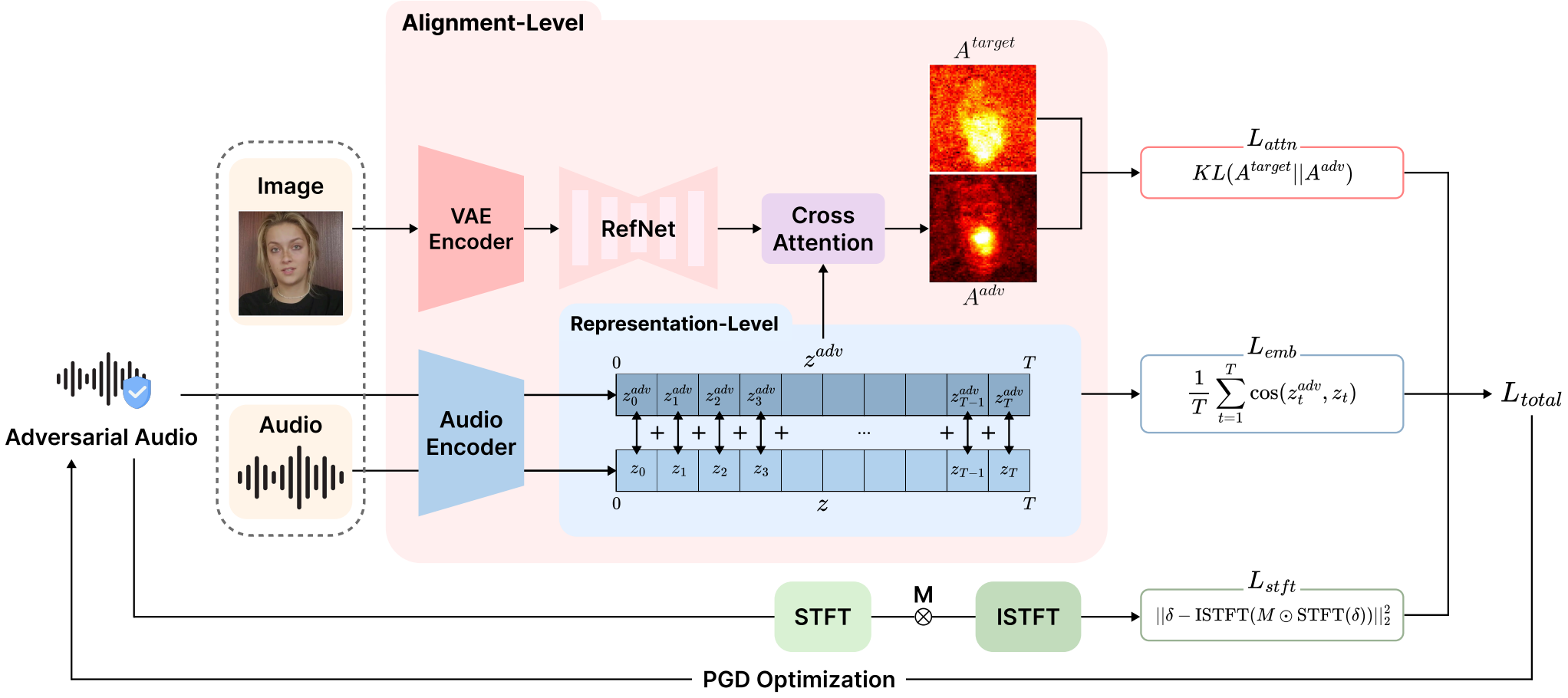}
    \end{minipage}
      \vspace{-7mm}
  \caption{\textbf{Overall Framework of MUTE.} The framework uncouples audio-visual alignment through a multi-level strategy, combining alignment-level disruption and representation-level degradation, while applying an STFT loss to ensure robustness.}
  \label{fig:framework}
  \vspace{-5mm}
\end{figure*}

We propose MUTE (Multi-level Alignment Uncoupling against Talking-Head Exploitation), an audio protection framework that degrades audio-visual synchronization quality in talking-head generation. As illustrated in Fig.\ref{fig:framework}, MUTE adopts a multi-level strategy to disrupt alignment through two complementary components: (1) \textbf{Alignment-level Disruption}, which directly perturbs the cross-modal alignment structure, and (2) \textbf{Representation-level Degradation}, which interferes with the extraction of reliable audio representations. These two components target different aspects of the alignment process and jointly amplify the overall disruption effectiveness.

At the alignment level, we perform \textbf{Alignment-level Disruption} (Section~\ref{sec:ali}) by directly perturbing the structural mechanism that governs audio-visual alignment. Specifically, we treat the attention map from a surrogate model as a proxy for the alignment structure and guide it toward a predefined target distribution. This directly disrupts the cross-modal correspondence between audio and visual features, weakening the alignment required for accurate lip synchronization.

Building upon this, we introduce \textbf{Representation-level Degradation} (Section~\ref{sec:rep}) by destabilizing the audio embedding extraction process. Since temporal audio representations drive frame-level facial motion, we decompose the embeddings along the temporal axis and apply degradation at each frame. By degrading the representations used for alignment, this step further propagates errors throughout the generation pipeline and amplifies synchronization disruption compared to applying either component alone.

Furthermore, to ensure \textbf{Robustness to Post-processing, Denoising and Adaptive Attack} (Section~\ref{sec:ada}), we design perturbations in the STFT domain by concentrating them in frequency regions less susceptible to removal. This allows MUTE to remain effective under transformations such as post-processing and denoising, thereby improving robustness in real-world. Additionally, to mitigate potential bypass via TTS-based resynthesis, we optionally incorporate a speaker-level objective to reduce identity consistency.

\subsection{Alignment-level Disruption}
\label{sec:ali}

To directly disrupt the audio-visual alignment structure, we leverage audio-conditioned visual cross-attention within the denoising UNet. Specifically, visual latent features encoded from a reference image $I^{ref}$ serve as queries, while audio embeddings extracted from the adversarial audio $x^{adv}$ serve as keys and values. This results in an attention map defined as $A^{adv} = \text{Attn}(Q(I^{ref}), K(x^{adv}), V(x^{adv}))$.

To disrupt alignment, we construct a target attention distribution by identifying spatial locations that are most sensitive to an $\epsilon$-bounded perturbation. To this end, we approximate the variation of attention using a Taylor expansion and derive a sensitivity score based on first- and second-order sensitivity:
$s(n) \propto \frac{|A'(n)| + \frac{\epsilon}{2}|A''(n)|}{A(n)}$, where $n$ indexes spatial locations in the attention map (e.g., 64 × 64 in our setting).
This formulation estimates the maximum achievable change under an $\epsilon$-bounded perturbation and prioritizes regions where attention can be effectively manipulated. Based on this score, we define the target attention distribution as $A^{target}(n) = \text{softmax}(s(n))$.

Finally, we define the alignment-level loss as a KL divergence between the target and adversarial attention distributions, given by
\begin{equation}
\mathcal{L}_{\text{attn}} = \mathrm{KL}(A^{target} \,\|\, A^{adv}) 
= \sum_n A^{target}(n) \log \frac{A^{target}(n)}{A^{adv}(n)}.
\end{equation}

The proposed Target Attention KL Loss treats the attention map as a spatial probability distribution and minimizes the discrepancy between the target and adversarial distributions. By penalizing cases where $A^{adv}$ fails to allocate sufficient probability mass to regions emphasized by $A^{target}$, it encourages attention redistribution toward those regions, effectively disrupting the underlying audio-visual alignment structure.

\subsection{Representation-level Degradation}
\label{sec:rep}

To degrade the audio representations used in talking-head generation, we leverage temporal audio embeddings extracted from the wav2vec2.0 \cite{wav2vec} encoder of the surrogate model, Hallo \cite{hallo}. Let $f_{\text{enc}}$ denote the audio encoder, and the extracted representation from the input audio $x$ is denoted as $z = f_{\text{enc}}(x) \in \mathbb{R}^{T \times L \times D}$, where $T$ denotes the number of temporal frames, $L$ the number of transformer layers, and $D$ the hidden dimension.

Since talking-head generation relies on temporal audio representations to drive frame-level facial motion, perturbing only a global embedding is insufficient to fully disrupt the generation process. Instead, we decompose the temporal embeddings into frame-wise representations and apply degradation at each time step, enabling more fine-grained and effective disruption of the generated motion.

Let $z_t$ denote the embedding at time step $t$, and $z_t^{adv}$ denote the corresponding embedding extracted from the adversarial audio $x^{adv}$. To realize the frame-wise degradation, we minimize the cosine similarity between corresponding embeddings, encouraging each temporal component to deviate from its original representation. The resulting loss is given by:
\begin{equation}
\mathcal{L}_{\text{emb}} =
\frac{1}{T}
\sum_{t=1}^{T}
\cos \left( z_t^{adv}, z_t \right).
\end{equation}

Compared to a global embedding objective that operates on the time-averaged representation, the proposed Frame-wise Embedding Cosine Loss directly penalizes each temporal component. This prevents local perturbations from being smoothed out through temporal aggregation and instead enforces cumulative degradation across the sequence, leading to more effective disruption of the generated motion.

\subsection{Robustness to Post-processing, Denoising and Adaptive Attack}
\label{sec:ada}

To improve robustness against perturbation removal, we design perturbations that remain effective under various real-world transformations. In particular, post-processing techniques and model-based denoisers tend to suppress low-energy noise components, making it crucial to prevent perturbations from being easily eliminated. To this end, we introduce an STFT loss that constrains the perturbation $\delta$ in the time-frequency domain as 
\begin{equation}
\mathcal{L}_{\text{stft}} = \left\| \delta - \text{ISTFT}\left(M \odot \text{STFT}(\delta)\right) \right\|_2^2, 
\end{equation}
encouraging the perturbation to reside in robust regions.

The mask $M$ is constructed from the STFT of clean audio, assigning a weight of 1 to high-energy time-frequency regions corresponding to the speech signal, while applying reduced weights (\ie less than 1) to low-energy or out-of-band components. This design biases the perturbation toward high-energy regions that are less likely to be removed by filtering or denoising processes.

As a result, the perturbations become spectrally aligned with the underlying speech signal and are less distinguishable as independent noise, making them more robust in practical scenarios.

\paragraph{TTS-based Resynthesis Robustness (Optional).}

In scenarios where an attacker is aware of the presence of perturbations, adversarial audio may be further processed to bypass the attack, for example, by extracting speaker identity and resynthesizing speech with arbitrary text. To mitigate such potential misuse, we optionally incorporate a speaker-level perturbation objective that can be combined with the main loss. This objective aims to reduce speaker identity consistency in the adversarial audio, making it difficult to preserve the original speaker characteristics under such transformations. 

To this end, we minimize the cosine similarity between speaker embeddings extracted from the original and adversarial audio, defined as $\mathcal{L}_{spk} = \cos(E_{spk}(x), E_{spk}(x^{adv}))$, encouraging the adversarial audio to deviate from the original speaker identity.
To improve robustness across different embedding spaces, we employ an ensemble of two speaker encoders, CAM++~\cite{campp} and a ResNet-based speaker encoder~\cite{eres2net}, and aggregate their losses. As a result, the adversarial audio becomes less consistent in speaker identity, making it more difficult for TTS-based resynthesis to preserve or reconstruct the original speaker characteristics.

\subsection{Final Objective and Optimization}

We adopt a two-phase optimization strategy to effectively perform alignment-level and representation-level disruption. Specifically, we partition the optimization process into two phases over a total of $K$ iterations. In the first phase, we focus on alignment-level disruption, while in the second phase, the optimization shifts to representation-level degradation. The objective at iteration $k$ is defined as:
\begin{equation}
\mathcal{L}_{\text{total}}^{(k)} =
\begin{cases}
\mathcal{L}_{\text{attn}} + \lambda \mathcal{L}_{\text{stft}}, & k \le \rho K \\
\mathcal{L}_{\text{emb}} + \lambda \mathcal{L}_{\text{stft}}, & k > \rho K
\end{cases},
\end{equation}
where $\rho \in (0,1)$ controls the proportion of iterations allocated to alignment-level optimization. The perturbation $\delta$ is optimized using PGD~\cite{pgd} under the loss at each phase.

This two-phase scheme separates objectives with different optimization directions, reducing gradient interference and improving optimization effectiveness. By first perturbing the alignment structure and then degrading audio representations, it achieves stronger misalignment. Furthermore, compared to joint optimization, this two-phase strategy reduces computational cost while allowing each phase to focus on its objective, resulting in a more efficient process. When the optional speaker-level objective is used, the final objective becomes $\mathcal{L}^{(k)} = \mathcal{L}_{\text{total}}^{(k)} + \lambda_{\text{spk}} \mathcal{L}_{\text{spk}}$.

\section{Experiments}
\label{sec:experiments}

\subsection{Experimental Setup}
\noindent\textbf{Datasets and Baselines.}
We use the TalkingHead-1KH~\cite{talkinghead1kh} and CelebV-HQ~\cite{celebvhq} datasets, which provide high-quality audio-visual samples with consistent speaker identity, for adversarial example generation and alignment evaluation. For adversarial example generation, we paired the input audio with a randomly selected reference image of a different identity. During evaluation, we instead used image–audio pairs with the same identity to assess the impact of adversarial perturbations under realistic conditions. For CelebV-HQ, we selected 50 high-quality video samples based on CLIP-IQA~\cite{clipiqa} scores. The corresponding audio was trimmed into 5-second segments. For TalkingHead-1KH, we follow the preprocessing protocol of Silencer~\cite{silencer}, using the same set of 50 samples. As baselines, we adopt representative audio adversarial attack methods: \emph{C\&W}~\cite{carlini}, \emph{Psychoacoustic}~\cite{psychoacoustic}, \emph{Qin et al.}~\cite{qin}, \emph{ARO}~\cite{aro}, \emph{SAGO}~\cite{sago} and \emph{CAF}~\cite{syncbreaker}. 

\noindent\textbf{Evaluation Metrics.}
We evaluate our method across three categories: \emph{audio quality}, \emph{audio-visual alignment}, and \emph{visual quality}. First, to assess the imperceptibility of adversarial perturbations added to the audio, we evaluate the protected audio quality. We report WER (\%) measured using Whisper~\cite{whisper} to assess speech intelligibility via ASR, along with STOI~\cite{stoi} for objective intelligibility and PESQ~\cite{pesq} for perceptual quality. Second, as the primary objective of our attack is to disrupt lip synchronization, we evaluate audio-visual alignment to measure how effectively the generated talking-head outputs are degraded. We use SyncNet~\cite{syncnet} to assess lip–audio synchronization and M-LMD~\cite{atvgnet} to quantify lip landmark distance. Finally, to measure secondary effects on the generated outputs, we evaluate visual quality as an auxiliary metric. We report FID~\cite{fid}, FVD~\cite{fvd}, and E-FID~\cite{hallo3}, where E-FID computes distributional distance over facial expression parameters.

\noindent\textbf{Implementation Details.}
We adopted Hallo~\cite{hallo} as the surrogate LDM-based talking-head model. All input portraits were resized to $512 \times 512$, and audio was sampled at 16 kHz. Adversarial perturbations were generated using PGD in the waveform space $[-1, 1]$ with an $L_\infty$ budget of $\epsilon=0.004$, a step size of $\alpha=0.0002$, and 200 iterations. For fair comparison, the perturbation budgets of baseline methods were adjusted to achieve comparable perceptual speech quality rather than using an identical numerical $\epsilon$ across all methods: $\epsilon=0.01$ for \emph{SAGO}~\cite{sago}, $\epsilon=0.006$ for \emph{Qin et al.}~\cite{qin}, and $\epsilon=0.004$ for MUTE and the remaining methods. Baseline methods were applied using the same wav2vec2.0~\cite{wav2vec} encoder used in Hallo~\cite{hallo}, under otherwise identical hyperparameter settings. We employed a two-phase optimization strategy with $\rho = 0.1$, where the first 10\% of iterations optimize the alignment-level objective and the remaining iterations optimize the representation-level objective. For $\mathcal{L}_{\text{emb}}$, the temporal audio embeddings are aligned to the 25 fps video frame rate, yielding 25 frame-wise embeddings per second. The STFT loss $\mathcal{L}_{\text{stft}}$ is weighted by $\lambda = 5$. We repeat the main experiments using three independent random seeds to account for stochasticity from diffusion sampling and PGD initialization. Results are reported as mean $\pm$ standard deviation across the three runs. All experiments were conducted on a single NVIDIA RTX 4090 GPU.

\begin{table}[t]
\centering
\caption{Quantitative comparison of MUTE against various audio adversarial attack baselines using TalkingHead-1KH and CelebV-HQ datasets.}
\label{tab:comparison}

\fontsize{6.5pt}{7.5pt}\selectfont %
\setlength{\tabcolsep}{0.3pt}       %
\renewcommand{\arraystretch}{0.95}

\renewcommand{\ms}[2]{%
  \begin{tabular}[c]{@{}c@{}}#1\\[-2.5pt]{\fontsize{4.5pt}{5.5pt}\selectfont $\pm$#2}\end{tabular}%
}

\begin{tabular}{l cc | cc cc || cc | cc cc}
\toprule
\multirow{3}{*}{Method}
& \multicolumn{6}{c}{TalkingHead-1KH~\cite{talkinghead1kh}}
& \multicolumn{6}{c}{CelebV-HQ~\cite{celebvhq}} \\

\cmidrule(lr){2-7}\cmidrule(lr){8-13}

& \multicolumn{2}{c}{Audio Quality}
& \multicolumn{2}{c}{A-V Align.}
& \multicolumn{2}{c}{Visual Quality}
& \multicolumn{2}{c}{Audio Quality}
& \multicolumn{2}{c}{A-V Align.}
& \multicolumn{2}{c}{Visual Quality} \\

\cmidrule(lr){2-3}\cmidrule(lr){4-5}\cmidrule(lr){6-7}
\cmidrule(lr){8-9}\cmidrule(lr){10-11}\cmidrule(lr){12-13}

& WER$\downarrow$
& STOI/PESQ$\uparrow$
& Sync$\downarrow$
& M-LMD$\uparrow$
& FID/FVD$\uparrow$
& E-FID$\uparrow$
& WER$\downarrow$
& STOI/PESQ$\uparrow$
& Sync$\downarrow$
& M-LMD$\uparrow$
& FID/FVD$\uparrow$
& E-FID$\uparrow$ \\

\midrule

C\&W~\cite{carlini}
& \ms{13.397}{0.488}
& \ms{0.922}{0.000} / \ms{2.322}{0.002}
& \ms{3.740}{0.016}
& \ms{1.096}{0.013}
& \ms{1.855}{0.083} / \ms{1.442}{0.024}
& \ms{3.646}{0.111}
& \ms{13.057}{0.210}
& \ms{0.948}{0.001} / \ms{2.702}{0.004}
& \ms{4.039}{0.032}
& \ms{1.108}{0.006}
& \ms{1.662}{0.151} / \ms{1.247}{0.030}
& \ms{3.508}{0.123}
\\

Psychoacoustic~\cite{psychoacoustic}
& \ms{12.140}{0.543}
& \ms{0.922}{0.001} / \ms{2.333}{0.003}
& \ms{3.787}{0.042}
& \ms{1.097}{0.005}
& \ms{1.914}{0.046} / \ms{1.427}{0.013}
& \ms{3.655}{0.110}
& \ms{11.643}{0.098}
& \ms{0.949}{0.001} / \ms{2.740}{0.002}
& \ms{3.926}{0.019}
& \ms{1.126}{0.011}
& \ms{1.676}{0.202} / \ms{1.276}{0.055}
& \ms{3.457}{0.191}
\\

Qin et al.~\cite{qin}
& \ms{15.649}{0.652}
& \ms{0.925}{0.000} / \ms{2.301}{0.002}
& \ms{4.468}{0.031}
& \ms{0.983}{0.003}
& \ms{1.825}{0.144} / \ms{1.329}{0.039}
& \ms{3.443}{0.076}
& \ms{13.870}{0.079}
& \ms{0.942}{0.000} / \ms{2.544}{0.005}
& \ms{4.543}{0.025}
& \ms{1.054}{0.005}
& \ms{1.479}{0.018} / \ms{1.182}{0.016}
& \ms{3.331}{0.037}
\\

ARO~\cite{aro}
& \ms{15.673}{0.139}
& \ms{0.903}{0.000} / \ms{2.518}{0.044}
& \ms{3.518}{0.044}
& \ms{1.155}{0.004}
& \ms{1.983}{0.065} / \ms{1.510}{0.011}
& \ms{3.662}{0.061}
& \ms{13.870}{0.495}
& \ms{0.931}{0.000} / \ms{2.689}{0.002}
& \ms{3.818}{0.044}
& \ms{1.176}{0.006}
& \ms{1.574}{0.109} / \ms{1.260}{0.007}
& \ms{3.457}{0.083}
\\

SAGO~\cite{sago}
& \ms{12.463}{0.140}
& \ms{0.930}{0.001} / \ms{2.515}{0.002}
& \ms{3.264}{0.106}
& \ms{1.200}{0.010}
& \ms{2.176}{0.178} / \ms{1.667}{0.145}
& \ms{4.115}{0.471}
& \ms{11.970}{0.173}
& \ms{0.956}{0.001} / \ms{2.912}{0.002}
& \ms{3.597}{0.038}
& \ms{1.226}{0.026}
& \ms{1.894}{0.017} / \ms{1.399}{0.061}
& \ms{3.963}{0.063}
\\

CAF~\cite{syncbreaker}
& \ms{12.143}{0.295}
& \ms{0.925}{0.000} / \ms{2.292}{0.002}
& \ms{2.173}{0.025}
& \ms{1.341}{0.016}
& \ms{2.928}{0.182} / \ms{2.268}{0.068}
& \ms{5.435}{0.078}
& \ms{14.624}{0.501}
& \ms{0.947}{0.001} / \ms{2.652}{0.001}
& \ms{2.570}{0.094}
& \ms{1.334}{0.003}
& \ms{2.732}{0.012} / \ms{1.709}{0.021}
& \ms{4.734}{0.087}
\\

\midrule

\textbf{MUTE}
& \ms{\textbf{11.681}}{0.330}
& \ms{\textbf{0.936}}{0.001} / \ms{\textbf{2.522}}{0.002}
& \ms{\textbf{1.332}}{0.012}
& \ms{\textbf{1.627}}{0.012}
& \ms{\textbf{3.183}}{0.036} / \ms{\textbf{2.363}}{0.049}
& \ms{\textbf{5.489}}{0.087}
& \ms{\textbf{10.467}}{0.406}
& \ms{\textbf{0.959}}{0.000} / \ms{\textbf{2.940}}{0.001}
& \ms{\textbf{2.152}}{0.040}
& \ms{\textbf{1.693}}{0.049}
& \ms{\textbf{2.756}}{0.048} / \ms{\textbf{1.790}}{0.042}
& \ms{\textbf{5.062}}{0.183}
\\

\midrule\midrule

GT
& 0.000
& 1.000 / 4.500
& 5.656
& 0.000
& 0.000 / 0.000
& 0.000
& 0.000
& 1.000 / 4.500
& 5.294
& 0.000
& 0.000 / 0.000
& 0.000
\\

\bottomrule
\end{tabular}
\end{table}

\begin{figure*}[t]
	\centering
  \includegraphics[width=1.0\linewidth]{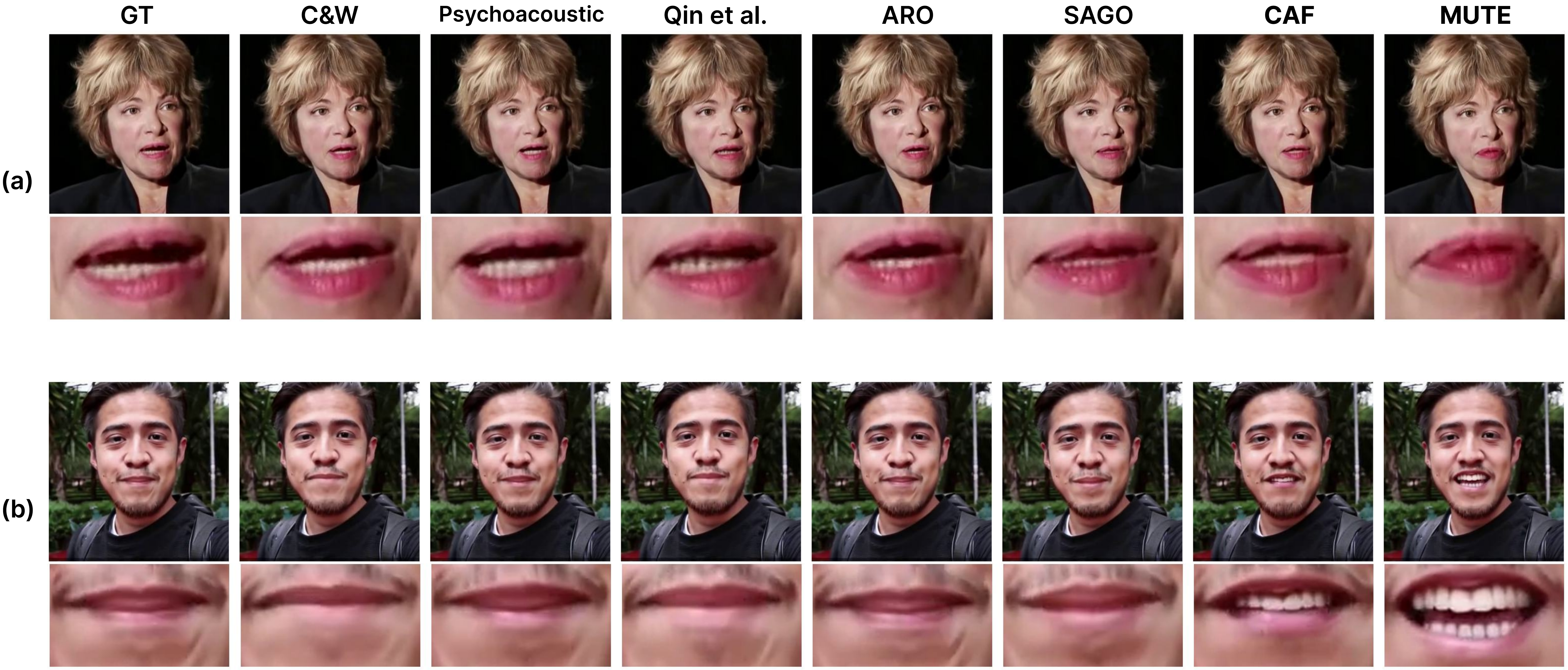}
      \vspace{-2mm}
  \caption{\textbf{Qualitative results on different datasets.} Visual comparison between baseline methods and MUTE on (a) TalkingHead-1KH dataset and (b) CelebV-HQ dataset.}
  \label{fig:qualitative}
    \vspace{-5mm}
\end{figure*}

\subsection{Evaluation on Voice Protection}

As summarized in Table~\ref{tab:comparison}, MUTE significantly outperforms baselines in disrupting audio-visual alignment. Our method achieves the lowest SyncNet scores (1.332 for TalkingHead-1KH and 2.152 for CelebV-HQ) and the highest M-LMD values (1.627 and 1.693), indicating stronger degradation of lip synchronization. Despite its strong protection capability, MUTE achieves the best perceptual audio quality among the compared methods, obtaining the lowest WER and highest STOI and PESQ scores while effectively disrupting facial dynamics. This demonstrates that our multi-level uncoupling strategy degrades alignment without compromising speech intelligibility. The qualitative effectiveness of MUTE is illustrated in Fig.\ref{fig:qualitative}, where our method induces the most significant disparity in facial dynamics and lip movements among all methods compared to ground truth and baselines.

\begin{table}
    \centering
    \vspace{-3mm}
    \caption{Performance comparison of combining Image-level Protection Methods with Voice-level Protection Methods.}
    \label{tab:silencer}
    \renewcommand{\arraystretch}{0.8}

    \renewcommand{\ms}[2]{%
      $#1\,{\scriptstyle\pm #2}$%
    }

    \resizebox{\linewidth}{!}{%
        \begin{tabular}{cc|ccccc}
        \toprule
        \multicolumn{1}{c}{Image Protection}
        & \multicolumn{1}{c|}{Voice Protection}
        & Sync$\downarrow$
        & M-LMD$\uparrow$
        & FID$\uparrow$
        & FVD$\uparrow$
        & E-FID$\uparrow$ \\
        \midrule

        MIS~\cite{syncbreaker}
        & -
        & \ms{3.039}{0.116}
        & \ms{1.988}{0.032}
        & \ms{142.692}{2.911}
        & \ms{23.009}{0.648}
        & \ms{162.260}{1.108}
        \\

        Silencer~\cite{silencer}
        & -
        & \ms{2.170}{0.014}
        & \ms{2.168}{0.013}
        & \ms{144.555}{1.254}
        & \ms{24.029}{0.513}
        & \ms{160.298}{3.172}
        \\
        \midrule

        MIS~\cite{syncbreaker}
        & CAF~\cite{syncbreaker}
        & \ms{1.212}{0.081}
        & \ms{2.157}{0.018}
        & \ms{145.166}{2.538}
        & \ms{24.526}{0.895}
        & \ms{163.048}{2.168}
        \\

        Silencer~\cite{silencer}
        & CAF~\cite{syncbreaker}
        & \ms{0.941}{0.025}
        & \ms{2.152}{0.007}
        & \ms{146.111}{0.495}
        & \ms{25.764}{0.079}
        & \ms{165.884}{0.282}
        \\

        \midrule

        MIS~\cite{syncbreaker}
        & MUTE
        & \ms{0.902}{0.055}
        & \ms{\textbf{2.261}}{0.046}
        & \ms{\textbf{147.141}}{2.480}
        & \ms{25.289}{0.456}
        & \ms{162.983}{0.583}
        \\

        Silencer~\cite{silencer}
        & MUTE
        & \ms{\textbf{0.760}}{0.004}
        & \ms{2.233}{0.015}
        & \ms{146.544}{0.221}
        & \ms{\textbf{25.843}}{0.087}
        & \ms{\textbf{168.829}}{1.553}
        \\

        \bottomrule
        \end{tabular}
    }
    \vspace{-3mm}
\end{table}

\noindent\textbf{Combination with Image Protection.}
Furthermore, MUTE demonstrates strong complementarity with image-level protection methods. As shown in Table~\ref{tab:silencer}, adding MUTE to either MIS~\cite{syncbreaker} or Silencer~\cite{silencer} substantially strengthens protection over image-only protection, markedly reducing the Sync score in both cases. Compared with the corresponding \emph{CAF}~\cite{syncbreaker} combinations, MUTE consistently achieves lower Sync scores and higher M-LMD scores, while inducing comparable or stronger visual-quality degradation. These results confirm that MUTE's audio-visual uncoupling synergizes effectively with image-domain perturbations to induce more severe alignment disruption.

\subsection{Performance on Black-Box Models}
We evaluate the transferability of MUTE under the transfer-based black-box setting, where adversarial perturbations are generated using the surrogate model (Hallo~\cite{hallo}) and applied to unseen talking-head models. Specifically, we evaluate on four target models: Hallo2~\cite{hallo2}, Audio2Head~\cite{audio2head}, SadTalker~\cite{sadtalker}, and Ditto~\cite{ditto}. Note that Hallo2 shares a similar architecture with the surrogate model, while the remaining models differ significantly in design. As shown in Table~\ref{tab:transferability}, MUTE achieves the lowest SyncNet and highest M-LMD across all target models, indicating strong degradation of audio-visual alignment, which in turn demonstrates its transferability to unseen talking-head architectures.

\begin{table*}[t]
\scriptsize
\centering
\caption{Transferability evaluation under black-box settings across talking-head models.}
\label{tab:transferability}
\renewcommand{\arraystretch}{0.85} %
\setlength{\tabcolsep}{9pt}        %
\renewcommand{\ms}[2]{%
  \begin{tabular}[c]{@{}c@{}}#1\\[-2.5pt]{\fontsize{4.5pt}{5.5pt}\selectfont $\pm$#2}\end{tabular}%
}

\begin{tabular}{l cc | cc | cc | cc}
\toprule
\multirow{2}{*}{Method} 
  & \multicolumn{2}{c}{Hallo2~\cite{hallo2}} 
  & \multicolumn{2}{c}{Audio2Head~\cite{audio2head}} 
  & \multicolumn{2}{c}{SadTalker~\cite{sadtalker}} 
  & \multicolumn{2}{c}{Ditto~\cite{ditto}} \\
\cmidrule(lr){2-3} \cmidrule(lr){4-5} \cmidrule(lr){6-7} \cmidrule(lr){8-9}
  & Sync$\downarrow$ & M-LMD$\uparrow$ 
  & Sync$\downarrow$ & M-LMD$\uparrow$ 
  & Sync$\downarrow$ & M-LMD$\uparrow$ 
  & Sync$\downarrow$ & M-LMD$\uparrow$ \\
\midrule
C\&W~\cite{carlini}
  & \ms{3.773}{0.053} & \ms{1.144}{0.018}
  & \ms{4.269}{0.014} & \ms{1.222}{0.044}
  & \ms{5.540}{0.013} & \ms{0.629}{0.006}
  & \ms{4.024}{0.019} & \ms{0.697}{0.004} \\

Psychoacoustic~\cite{psychoacoustic}
  & \ms{3.747}{0.034} & \ms{1.137}{0.011}
  & \ms{4.241}{0.018} & \ms{1.188}{0.016}
  & \ms{5.503}{0.056} & \ms{0.630}{0.004}
  & \ms{4.002}{0.008} & \ms{0.697}{0.001} \\

Qin et al.~\cite{qin}
  & \ms{4.535}{0.017} & \ms{1.027}{0.018}
  & \ms{4.510}{0.029} & \ms{1.104}{0.013}
  & \ms{5.702}{0.067} & \ms{0.583}{0.002}
  & \ms{4.182}{0.038} & \ms{0.678}{0.004} \\

ARO~\cite{aro}
  & \ms{3.546}{0.047} & \ms{1.192}{0.011}
  & \ms{4.182}{0.025} & \ms{1.203}{0.005}
  & \ms{5.297}{0.027} & \ms{0.658}{0.002}
  & \ms{3.803}{0.039} & \ms{0.781}{0.005} \\

SAGO~\cite{sago}
  & \ms{3.278}{0.010} & \ms{1.219}{0.014}
  & \ms{4.335}{0.022} & \ms{1.133}{0.004}
  & \ms{5.418}{0.018} & \ms{0.563}{0.006}
  & \ms{3.959}{0.009} & \ms{0.704}{0.006} \\

CAF~\cite{syncbreaker}
  & \ms{2.280}{0.042} & \ms{1.373}{0.016}
  & \ms{4.005}{0.010} & \ms{1.246}{0.007}
  & \ms{5.425}{0.012} & \ms{0.642}{0.003}
  & \ms{3.994}{0.024} & \ms{0.703}{0.004} \\
\midrule
\textbf{MUTE}
  & \ms{\textbf{1.397}}{0.057} & \ms{\textbf{1.679}}{0.013}
  & \ms{\textbf{3.627}}{0.036} & \ms{\textbf{1.299}}{0.023}
  & \ms{\textbf{4.703}}{0.059} & \ms{\textbf{0.864}}{0.008}
  & \ms{\textbf{3.457}}{0.037} & \ms{\textbf{0.884}}{0.006} \\
\midrule
\midrule
GT & 5.688 & 0.000 & 5.314 & 0.000 & 6.389 & 0.000 & 4.778 & 0.000 \\
\bottomrule
\end{tabular}
\vspace{-5mm}
\end{table*}

\begin{table*}[t]
\scriptsize %
\centering
\caption{Robustness evaluation under post-processing, denoising, and adaptive attack.}
\label{tab:robustness}
\renewcommand{\arraystretch}{0.85} %
\setlength{\tabcolsep}{4pt}        %
\renewcommand{\ms}[2]{%
  \begin{tabular}[c]{@{}c@{}}#1\\[-2.5pt]{\fontsize{4.5pt}{5.5pt}\selectfont $\pm$#2}\end{tabular}%
}

\begin{tabular}{l cc | cc | cc | cc | cc}
\toprule
\multirow{2}{*}{Method} 
  & \multicolumn{2}{c}{Down/Upsampling~\cite{waveguard}} 
  & \multicolumn{2}{c}{Spectral Filtering~\cite{waveguard}} 
  & \multicolumn{2}{c}{Demucs~\cite{demucs}} 
  & \multicolumn{2}{c}{WavePurifier~\cite{wavepurifier}} 
  & \multicolumn{2}{c}{Adaptive Attack~\cite{tramer}} \\
\cmidrule(lr){2-3} \cmidrule(lr){4-5} \cmidrule(lr){6-7} \cmidrule(lr){8-9} \cmidrule(lr){10-11}
  & Sync$\downarrow$ & M-LMD$\uparrow$ 
  & Sync$\downarrow$ & M-LMD$\uparrow$ 
  & Sync$\downarrow$ & M-LMD$\uparrow$ 
  & Sync$\downarrow$ & M-LMD$\uparrow$ 
  & Sync$\downarrow$ & M-LMD$\uparrow$ \\
\midrule
C\&W~\cite{carlini}
  & \ms{3.791}{0.052} & \ms{0.971}{0.002}
  & \ms{3.884}{0.026} & \ms{0.987}{0.010}
  & \ms{3.968}{0.033} & \ms{1.031}{0.009}
  & \ms{4.071}{0.016} & \ms{0.940}{0.013}
  & \ms{4.878}{0.021} & \ms{0.557}{0.016} \\

Psychoacoustic~\cite{psychoacoustic}
  & \ms{3.825}{0.028} & \ms{0.979}{0.019}
  & \ms{3.878}{0.013} & \ms{1.004}{0.003}
  & \ms{3.953}{0.055} & \ms{1.042}{0.001}
  & \ms{4.059}{0.009} & \ms{0.935}{0.001}
  & \ms{4.840}{0.034} & \ms{0.555}{0.015} \\

Qin et al.~\cite{qin}
  & \ms{4.032}{0.044} & \ms{0.955}{0.007}
  & \ms{4.346}{0.052} & \ms{0.950}{0.022}
  & \ms{4.569}{0.041} & \ms{0.978}{0.011}
  & \ms{3.974}{0.123} & \ms{0.952}{0.024}
  & \ms{4.926}{0.040} & \ms{0.564}{0.009} \\

ARO~\cite{aro}
  & \ms{3.607}{0.033} & \ms{1.069}{0.008}
  & \ms{3.647}{0.022} & \ms{1.074}{0.009}
  & \ms{3.612}{0.040} & \ms{1.132}{0.007}
  & \ms{4.059}{0.030} & \ms{0.985}{0.011}
  & \ms{4.778}{0.032} & \ms{0.555}{0.004} \\

SAGO~\cite{sago}
  & \ms{3.805}{0.088} & \ms{1.017}{0.007}
  & \ms{3.801}{0.045} & \ms{1.087}{0.041}
  & \ms{3.407}{0.029} & \ms{1.141}{0.018}
  & \ms{4.101}{0.023} & \ms{0.922}{0.009}
  & \ms{4.912}{0.022} & \ms{0.562}{0.006} \\

CAF~\cite{syncbreaker}
  & \ms{3.533}{0.048} & \ms{1.089}{0.011}
  & \ms{3.063}{0.078} & \ms{1.178}{0.020}
  & \ms{2.566}{0.027} & \ms{1.259}{0.011}
  & \ms{3.940}{0.021} & \ms{0.991}{0.024}
  & \ms{4.810}{0.016} & \ms{0.563}{0.016} \\
\midrule
\textbf{MUTE}
  & \ms{\textbf{2.914}}{0.025} & \ms{\textbf{1.254}}{0.028}
  & \ms{\textbf{2.251}}{0.068} & \ms{\textbf{1.410}}{0.025}
  & \ms{\textbf{1.534}}{0.010} & \ms{\textbf{1.593}}{0.003}
  & \ms{\textbf{3.494}}{0.056} & \ms{\textbf{1.217}}{0.008}
  & \ms{\textbf{4.269}}{0.044} & \ms{\textbf{0.655}}{0.010} \\
\bottomrule
\end{tabular}
\end{table*}

\subsection{Robustness to Post-processing, Denoising and Adaptive Attack}

To evaluate the robustness of MUTE under realistic and adaptive attack settings, we consider post-processing, denoising, and optimization-based adaptive attack. For post-processing, we follow WaveGuard~\cite{waveguard} and apply downsampling–upsampling (16 kHz $\rightarrow$ 8 kHz $\rightarrow$ 16 kHz) and adaptive spectral filtering based on the median spectral centroid. For denoising, we employ Demucs~\cite{demucs} and WavePurifier~\cite{wavepurifier}, representing source separation and diffusion-based purification, respectively. Following the adaptive-attack methodology of~\cite{tramer}, we consider an adversary who is aware of the protection and optimizes a purification of the protected audio. For MUTE, the purification smooths STFT magnitudes within the 80--7600\,Hz band targeted by our STFT loss, whereas for baselines, which do not employ a spectral mask, it is applied over the full band. The non-differentiable masking step is handled with BPDA~\cite{bpda}. As shown in Table~\ref{tab:robustness}, MUTE achieves the lowest Sync and highest M-LMD among all methods under every setting. It remains effective under model-based denoising, and although adaptive attack reduces protection efficacy, MUTE still outperforms all baselines.

\noindent\textbf{Robustness against TTS-based Resynthesis.}
\label{sec:tts_robustness}

\begin{table}[t]
    \centering
    \caption{Analysis of the optional speaker-level objective across different $\lambda_{spk}$.}
    \vspace{-3mm}
    \label{tab:adaptive_attack}
    \renewcommand{\arraystretch}{0.9}

    \renewcommand{\ms}[2]{%
      $#1\,{\scriptstyle\pm #2}$%
    }

    \setlength{\tabcolsep}{2pt}
    \resizebox{\linewidth}{!}{%
    \begin{tabular}{l cc | cc | cc | cc}
    \toprule

    \multirow{2}{*}{\begin{tabular}[c]{@{}l@{}}Reference\\ Condition\end{tabular}}
    & \multicolumn{2}{c}{CosyVoice~\cite{cosyvoice}}
    & \multicolumn{2}{c}{XTTS-v2~\cite{xtts}}
    & \multicolumn{2}{c}{Chatterbox~\cite{chatterbox}}
    & \multirow{2}{*}{Sync$\downarrow$}
    & \multirow{2}{*}{M-LMD$\uparrow$} \\

    \cmidrule(lr){2-3}
    \cmidrule(lr){4-5}
    \cmidrule(lr){6-7}

    & Sim$\downarrow$ & DSR$\uparrow$
    & Sim$\downarrow$ & DSR$\uparrow$
    & Sim$\downarrow$ & DSR$\uparrow$
    & & \\

    \midrule

    Unprotected
      & \ms{0.488}{0.003} & \ms{2.53}{0.46}
      & \ms{0.469}{0.002} & \ms{2.80}{0.40}
      & \ms{0.574}{0.008} & \ms{4.53}{1.01}
      & 5.656 & 0.000 \\

    \midrule

    $0$
      & \ms{0.368}{0.005} & \ms{14.67}{1.89}
      & \ms{0.392}{0.002} & \ms{14.93}{1.67}
      & \ms{0.449}{0.009} & \ms{15.07}{2.66}
      & \ms{1.332}{0.012} & \ms{1.627}{0.012} \\

    $0.01$
      & \ms{0.200}{0.007} & \ms{67.73}{2.60}
      & \ms{-0.102}{0.005} & \ms{99.33}{0.83}
      & \ms{0.280}{0.005} & \ms{43.87}{0.23}
      & \ms{1.496}{0.067} & \ms{1.541}{0.012} \\

    $0.05$
      & \ms{0.180}{0.008} & \ms{71.47}{1.01}
      & \ms{-0.120}{0.002} & \ms{99.47}{0.46}
      & \ms{0.248}{0.005} & \ms{50.40}{2.62}
      & \ms{1.918}{0.053} & \ms{1.465}{0.008} \\

    $0.10$
      & \ms{0.170}{0.003} & \ms{74.13}{1.67}
      & \ms{-0.120}{0.004} & \ms{99.73}{0.23}
      & \ms{0.239}{0.002} & \ms{52.67}{2.89}
      & \ms{2.218}{0.042} & \ms{1.415}{0.009} \\

    \bottomrule
    \end{tabular}%
    }

    \vspace{-5mm}
\end{table}

We evaluate the robustness of MUTE against TTS-based resynthesis attacks, where an adversary reconstructs speech using protected audio as the speaker reference. In the white-box setting, we use CosyVoice~\cite{cosyvoice} and Coqui XTTS-v2~\cite{xtts}, while in the black-box setting we use Chatterbox~\cite{chatterbox}. We compute speaker similarity (Sim) using an ECAPA-TDNN-based~\cite{ecapatdnn} speaker verification model and define the defense success rate (DSR) as the proportion of synthesized samples classified as different speakers from the corresponding original audio, using a cosine similarity threshold of 0.25 or lower. As shown in Table~\ref{tab:adaptive_attack}, unprotected audio yields higher Sim and substantially lower DSR, indicating that the original speaker characteristics are largely preserved under TTS resynthesis. Increasing $\lambda_{\text{spk}}$ improves DSR at the cost of reduced alignment-disruption effectiveness, revealing a trade-off between reducing speaker identity consistency and disrupting audio-visual alignment. Notably, with $\lambda_{\text{spk}}=0.01$, MUTE achieves over 43\% DSR across all TTS models while maintaining strong alignment-disruption performance, with SyncNet and M-LMD scores of 1.496 and 1.541, respectively. These results demonstrate that the optional speaker-level objective effectively reduces speaker identity consistency under TTS resynthesis while preserving the primary goal of disrupting audio-visual alignment.

\subsection{User Study}
\begin{wrapfigure}{r}{0.42\columnwidth}
    \centering
    \vspace{-4mm}
    \includegraphics[width=\linewidth]{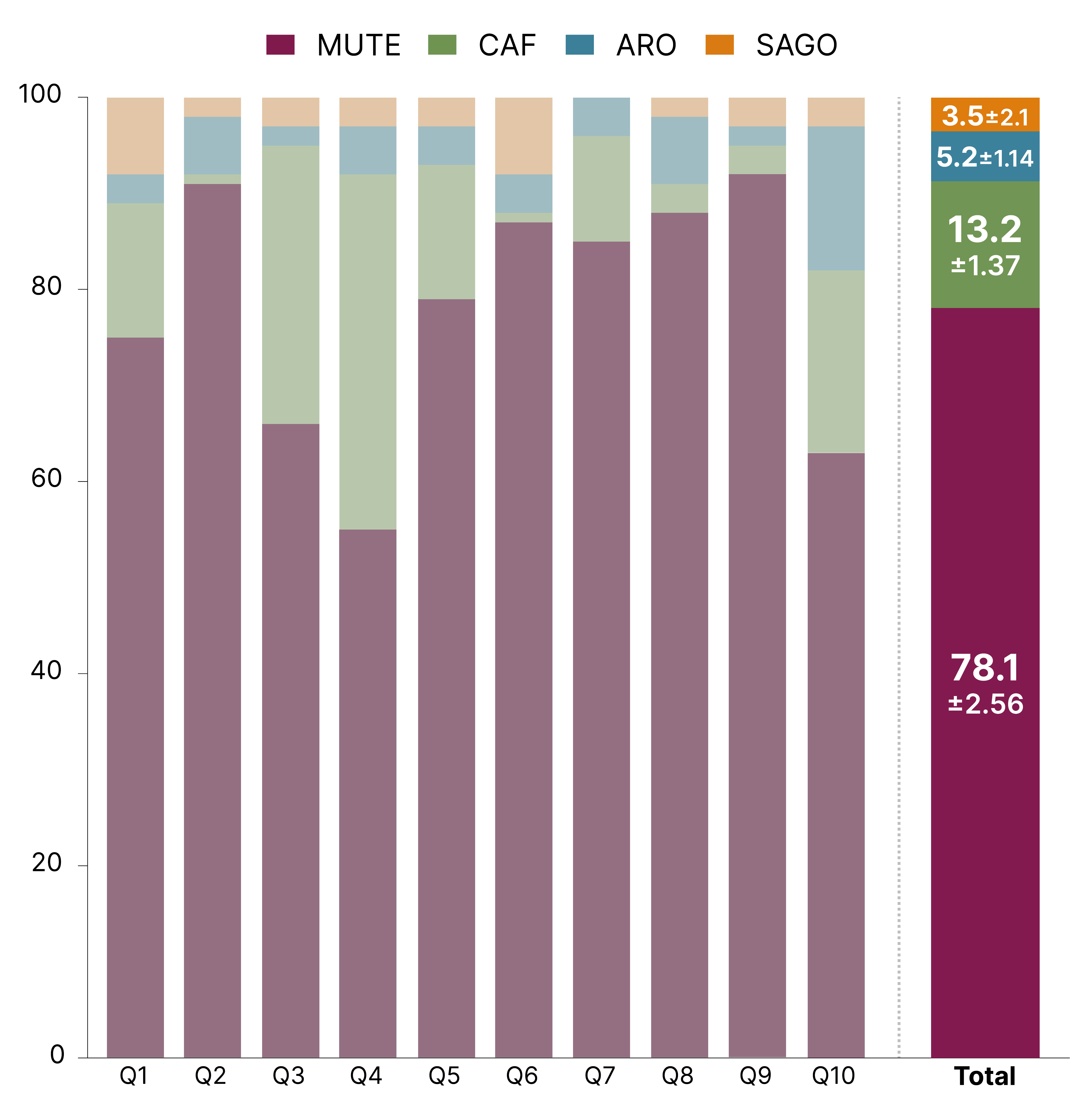}
    \vspace{-6mm}
    \caption{User study results.}
    \label{fig:user_study}
    \vspace{-3mm}
\end{wrapfigure}

While quantitative metrics such as SyncNet and M-LMD measure audio-visual alignment, they may not fully capture human perception of lip synchronization quality. To complement these metrics, we conduct a user study to evaluate whether the degradation introduced by MUTE is perceptually noticeable. We compare MUTE with three strong baseline methods, \emph{ARO}~\cite{aro}, \emph{SAGO}~\cite{sago}, and \emph{CAF}~\cite{syncbreaker}, using 10 video samples. For each sample, participants are presented with videos generated by different methods and asked to select the one that exhibits the most noticeable lip-sync mismatch. A total of 100 participants take part in the study, resulting in 1,000 responses. As shown in Fig.\ref{fig:user_study}, MUTE is selected as the most misaligned method in $78.1\% \pm 2.56\%$ (95\% CI) of cases, significantly outperforming all baselines. This demonstrates that the degradation introduced by MUTE is not only measurable by objective metrics but also clearly perceivable to human observers.

\subsection{Ablation Study}
\begin{table*}[t]
\centering
\setlength{\tabcolsep}{2.5pt}
\fontsize{6.5pt}{8pt}\selectfont %

\begin{minipage}[t]{0.32\textwidth}
    \centering
    \caption{Effect of ratio $\rho$.}
    \label{tab:kl_emb_ratio}
    \renewcommand{\arraystretch}{0.95}
    \setlength{\aboverulesep}{0pt}
    \setlength{\belowrulesep}{0pt}
    
    \begin{tabular}{c|c|c|c|c}
    \toprule
    $\rho$ & time$\downarrow$ & Hallo & Audio2Head & SadTalker \\
    \midrule
    1.0  & 78.4 & 3.920 & 3.958 & 5.048 \\
    \midrule
    0.2  & 20.1 & 1.419 & 3.798 & 4.898 \\
    \rowcolor{gray!10} 0.1 & 12.3 & \textbf{1.332} & \textbf{3.627} & \textbf{4.703} \\
    0.05 & 9.2  & 1.470 & 3.799 & 4.899 \\
    \midrule
    0.0  & 4.8  & 1.427 & 3.904 & 5.063 \\
    \bottomrule
    \end{tabular}
\end{minipage}
\hfill
\begin{minipage}[t]{0.31\textwidth}
    \centering
    \caption{Effect of weight $\lambda$.}
    \label{tab:robustness_weight}
    \renewcommand{\arraystretch}{0.95}
    \setlength{\aboverulesep}{0pt}
    \setlength{\belowrulesep}{0pt}
    \setlength{\tabcolsep}{6pt}
    
    \begin{tabular}{c|c|c|c}
    \toprule
    $\lambda$ & Clean & Down/Up & Demucs \\
    \midrule
    0   & 1.492 & 3.178 & 1.816 \\
    \rowcolor{gray!10} 5 & \textbf{1.332} & \textbf{2.914} & 1.534 \\
    10  & 1.429 & 2.933 & \textbf{1.494} \\
    20  & 1.451 & 2.971 & 1.564 \\
    50  & 1.474 & 3.024 & 1.675 \\
    \bottomrule
    \end{tabular}
\end{minipage}
\hfill
\begin{minipage}[t]{0.33\textwidth}
    \centering
    \caption{Effect of budget $\epsilon$.}
    \label{tab:epsilon_budget}
    \renewcommand{\arraystretch}{0.95}
    \setlength{\aboverulesep}{0pt}
    \setlength{\belowrulesep}{0pt}
    
    \begin{tabular}{c|c|c|c}
    \toprule
    $\epsilon$ & WER$\downarrow$(\%) & STOI$\uparrow$ / PESQ$\uparrow$ & Sync$\downarrow$ \\
    \midrule
    0.002 & \textbf{10.35} & \textbf{0.948} / \textbf{2.949} & 1.898 \\
    \rowcolor{gray!10} 0.004 & 11.68 & 0.936 / 2.522 & 1.332 \\
    0.006 & 12.25 & 0.907 / 2.298 & 1.200 \\
    0.008 & 12.89 & 0.891 / 2.108 & 1.122 \\
    0.010 & 13.30 & 0.885 / 1.945 & \textbf{1.027} \\
    \bottomrule
    \end{tabular}
\end{minipage}
\vspace{-5mm}
\end{table*}

\noindent\textbf{Multi-level Optimization Strategy.}

We analyze the effect of the two-phase strategy by varying $\rho \in \{1, 0.2, 0.1, 0.05, 0\}$, where $\rho = 1$ and $\rho = 0$ correspond to using only the alignment-level and representation-level objectives, respectively. As shown in Table~\ref{tab:kl_emb_ratio}, both objectives effectively disrupt audio-visual alignment. However, the representation-level objective achieves stronger white-box degradation, while the alignment-level objective shows better black-box transferability, revealing a trade-off between primary performance and cross-model generalization. The representation-level objective is also more computationally efficient. Balancing these factors, we adopt $\rho = 0.1$, which yields robust performance across all settings.

\noindent\textbf{STFT Loss Weight.}
We analyze the effect of the STFT loss by varying its weight $\lambda \in \{0, 5, 10, 20, 50\}$, where $\lambda = 0$ corresponds to optimizing only the alignment-level and representation-level objectives. We evaluate performance using SyncNet and M-LMD to measure audio-visual alignment. As shown in Table~\ref{tab:robustness_weight}, adding $\mathcal{L}_{\text{stft}}$ not only enhances the primary attack performance by further degrading lip synchronization but also improves robustness. Considering both alignment degradation and robustness, we select $\lambda = 5$ as it provides the best balance between performance and stability.

\noindent\textbf{Perturbation Budget.}
We analyze the trade-off between the perturbation budget and protection performance over $\epsilon \in \{0.002, 0.004, 0.006, 0.008, 0.010\}$. As $\epsilon$ increases, the audio-visual alignment disruption becomes progressively stronger, while perceptual audio quality gradually decreases. We identify $\epsilon=0.004$ as a balanced operating point, providing strong alignment disruption while maintaining high audio quality.

\section{Conclusions}
\label{sec:conclusions}

We propose MUTE, a multi-level audio-based protection framework that uncouples audio-visual alignment in talking-head generation. By combining Alignment-level Disruption and Representation-level Degradation, MUTE effectively degrades lip synchronization, which in turn enables effective transferability across unseen talking-head models. Furthermore, MUTE maintains robustness under real-world transformations, including post-processing, denoising, adaptive attacks and TTS-based resynthesis, while preserving perceptual audio quality. As an audio-only method, MUTE achieves strong protection on its own and can be further combined with image-based approaches for enhanced multimodal protection. These results demonstrate that targeting audio-visual alignment is an effective strategy for protecting against talking-head exploitation, providing a practical and extensible direction for multimodal data protection. Limitations and future work of MUTE are discussed in Appendix.
\section*{Acknowledgement}
This work was supported in part by Korea Internet \& Security Agency(KISA) grant funded by the Korea government(PIPC), Development of Robust Real-time Deepfake Disruption Technology via Adversarial Noise Injection for Privacy Protection, under Grant RS-2026-25530576; and in part by the Institute of Information \& communications Technology Planning \& Evaluation (IITP) grant funded by the Korea government(MSIT) (RS-2021-II211341).

\clearpage
\bibliography{main}

\begin{thebibliography}{49}
\providecommand{\natexlab}[1]{#1}
\providecommand{\url}[1]{\texttt{#1}}
\expandafter\ifx\csname urlstyle\endcsname\relax
  \providecommand{\doi}[1]{doi: #1}\else
  \providecommand{\doi}{doi: \begingroup \urlstyle{rm}\Url}\fi

\bibitem[Zhang et~al.(2023)Zhang, Cun, Wang, Zhang, Shen, Guo, Shan, and
  Wang]{sadtalker}
Wenxuan Zhang, Xiaodong Cun, Xuan Wang, Yong Zhang, Xi~Shen, Yu~Guo, Ying Shan,
  and Fei Wang.
\newblock {SadTalker}: Learning realistic {3D} motion coefficients for stylized
  audio-driven single image talking face animation.
\newblock In \emph{CVPR}, 2023.

\bibitem[Guo et~al.(2021)Guo, Chen, Liang, Liu, Bao, and Zhang]{adnerf}
Yudong Guo, Keyu Chen, Sen Liang, Yong-Jin Liu, Hujun Bao, and Juyong Zhang.
\newblock {AD-NeRF}: Audio driven neural radiance fields for talking head
  synthesis.
\newblock In \emph{ICCV}, 2021.

\bibitem[Fan et~al.(2022)Fan, Lin, Saito, Wang, and Komura]{faceformer}
Yingruo Fan, Zhaojiang Lin, Jun Saito, Wenping Wang, and Taku Komura.
\newblock {FaceFormer}: Speech-driven {3D} facial animation with transformers.
\newblock In \emph{CVPR}, 2022.

\bibitem[Wang et~al.(2021{\natexlab{a}})Wang, Li, Ding, Fan, and
  Yu]{audio2head}
Suzhen Wang, Lincheng Li, Yu~Ding, Changjie Fan, and Xin Yu.
\newblock {Audio2Head}: Audio-driven one-shot talking-head generation with
  natural head motion.
\newblock In \emph{ICCV}, 2021{\natexlab{a}}.

\bibitem[Xu et~al.(2024)Xu, Li, Su, Shang, Zhang, Liu, Wang, Yao, and
  Zhu]{hallo}
Mingwang Xu, Hui Li, Qingkun Su, Hanlin Shang, Liwei Zhang, Ce~Liu, Jingdong
  Wang, Yao Yao, and Siyu Zhu.
\newblock {Hallo}: Hierarchical audio-driven visual synthesis for portrait
  image animation.
\newblock \emph{arXiv preprint arXiv:2406.08801}, 2024.

\bibitem[Cui et~al.(2025{\natexlab{a}})Cui, Li, Yao, Zhu, Shang, Cheng, Zhou,
  Zhu, and Wang]{hallo2}
Jiahao Cui, Hui Li, Yao Yao, Hao Zhu, Hanlin Shang, Kaihui Cheng, Hang Zhou,
  Siyu Zhu, and Jingdong Wang.
\newblock {Hallo2}: Long-duration and high-resolution audio-driven portrait
  image animation.
\newblock In \emph{ICLR}, 2025{\natexlab{a}}.

\bibitem[Cui et~al.(2025{\natexlab{b}})Cui, Li, Zhan, Shang, Cheng, Ma, Mu,
  Zhou, Wang, and Zhu]{hallo3}
Jiahao Cui, Hui Li, Yun Zhan, Hanlin Shang, Kaihui Cheng, Yuqi Ma, Shan Mu,
  Hang Zhou, Jingdong Wang, and Siyu Zhu.
\newblock {Hallo3}: Highly dynamic and realistic portrait image animation with
  video diffusion transformer.
\newblock In \emph{CVPR}, 2025{\natexlab{b}}.

\bibitem[Ye et~al.(2023)Ye, He, Jiang, Huang, Huang, Liu, Ren, Yin, Ma, and
  Zhao]{genefacepp}
Zhenhui Ye, Jinzheng He, Ziyue Jiang, Rongjie Huang, Jiawei Huang, Jinglin Liu,
  Yi~Ren, Xiang Yin, Zejun Ma, and Zhou Zhao.
\newblock {GeneFace++}: Generalized and stable real-time audio-driven {3D}
  talking face generation.
\newblock \emph{arXiv preprint arXiv:2305.00787}, 2023.

\bibitem[Li et~al.(2025)Li, Zheng, Yang, Chen, and Yang]{ditto}
Tianqi Li, Ruobing Zheng, Minghui Yang, Jingdong Chen, and Ming Yang.
\newblock {Ditto}: Motion-space diffusion for controllable realtime talking
  head synthesis.
\newblock In \emph{ACM MM}, 2025.

\bibitem[Prajwal et~al.(2020)Prajwal, Mukhopadhyay, Namboodiri, and
  Jawahar]{wav2lip}
K~R Prajwal, Rudrabha Mukhopadhyay, Vinay~P Namboodiri, and C~V Jawahar.
\newblock A lip sync expert is all you need for speech to lip generation in the
  wild.
\newblock In \emph{ACM MM}, 2020.

\bibitem[Zhou et~al.(2020)Zhou, Han, Shechtman, Echevarria, Kalogerakis, and
  Li]{makeittalk}
Yang Zhou, Xintong Han, Eli Shechtman, Jose Echevarria, Evangelos Kalogerakis,
  and Dingzeyu Li.
\newblock {MakeItTalk}: Speaker-aware talking-head animation.
\newblock \emph{ACM TOG}, 39\penalty0 (6):\penalty0 1--15, 2020.

\bibitem[Jeong et~al.(2025)]{faceshield}
Jaehwan Jeong et~al.
\newblock {FaceShield}: Defending facial image against deepfake threats.
\newblock In \emph{ICCV}, 2025.

\bibitem[Wang et~al.(2025)]{nullswap}
Tianyi Wang et~al.
\newblock {NullSwap}: Proactive identity cloaking against deepfake face
  swapping.
\newblock In \emph{ICCV}, 2025.

\bibitem[Zhang et~al.(2024)]{dualdefense}
Yunming Zhang et~al.
\newblock Dual defense: Adversarial, traceable, and invisible robust
  watermarking against face swapping.
\newblock \emph{IEEE Trans. Inform. Forensics Secur.}, 19:\penalty0 4628--4641,
  2024.

\bibitem[Yu et~al.(2023)Yu, Zhai, and Zhang]{antifake}
Zhiyuan Yu, Shixuan Zhai, and Ning Zhang.
\newblock {AntiFake}: Using adversarial audio to prevent unauthorized speech
  synthesis.
\newblock In \emph{ACM CCS}, 2023.

\bibitem[Kim et~al.(2026)]{roco}
Seungmin Kim et~al.
\newblock {RoCo}: Robust code for fast and effective proactive defense against
  voice cloning attack.
\newblock In \emph{ICASSP}, 2026.

\bibitem[Zhang et~al.(2025)Zhang, Wang, Mi, Wu, Gao, Cao, Ye, Xue, and
  Hao]{e2evguard}
Zhisheng Zhang, Derui Wang, Yifan Mi, Zhiyong Wu, Jie Gao, Yuxin Cao, Kai Ye,
  Minhui Xue, and Jie Hao.
\newblock {E2E-VGuard}: Adversarial prevention for production {LLM}-based
  end-to-end speech synthesis.
\newblock In \emph{NeurIPS}, 2025.

\bibitem[Gan et~al.(2025)Gan, Miao, Wang, and Yang]{silencer}
Yuan Gan, Jiaxu Miao, Yunze Wang, and Yi~Yang.
\newblock Silence is golden: Leveraging adversarial examples to nullify audio
  control in ldm-based talking-head generation.
\newblock In \emph{CVPR}, 2025.

\bibitem[Zhang et~al.(2026)Zhang, Shi, Zhao, Chen, Cheng, Xu, Xu, and
  Liao]{syncbreaker}
Wenli Zhang, Xianglong Shi, Sirui Zhao, Xinqi Chen, Guo Cheng, Yifan Xu, Tong
  Xu, and Yong Liao.
\newblock {SyncBreaker}: Stage-aware multimodal adversarial attacks on
  audio-driven talking head generation.
\newblock \emph{arXiv preprint arXiv:2604.08405}, 2026.

\bibitem[Carlini and Wagner(2018)]{carlini}
Nicholas Carlini and David Wagner.
\newblock Audio adversarial examples: Targeted attacks on speech-to-text.
\newblock In \emph{IEEE SPW}, 2018.

\bibitem[Sch{\"o}nherr et~al.(2018)Sch{\"o}nherr, Kohls, Zeiler, Holz, and
  Kolossa]{psychoacoustic}
Lea Sch{\"o}nherr, Katharina Kohls, Steffen Zeiler, Thorsten Holz, and Dorothea
  Kolossa.
\newblock Adversarial attacks against automatic speech recognition systems via
  psychoacoustic hiding.
\newblock In \emph{NDSS}, 2018.

\bibitem[Qin et~al.(2019)Qin, Carlini, et~al.]{qin}
Yao Qin, Nicholas Carlini, et~al.
\newblock Imperceptible, robust, and targeted adversarial examples for
  automatic speech recognition.
\newblock In \emph{ICML}, 2019.

\bibitem[Jin et~al.(2025)Jin, Su, Wang, Ye, and Hao]{aro}
Weifei Jin, Junjie Su, Hejia Wang, Yulin Ye, and Jie Hao.
\newblock Boosting the transferability of audio adversarial examples with
  acoustic representation optimization.
\newblock In \emph{ICME}, 2025.

\bibitem[Gao et~al.(2024)Gao, Li, Chen, Liu, and Li]{sago}
Xiaoxue Gao, Zexin Li, Yiming Chen, Cong Liu, and Haizhou Li.
\newblock Transferable adversarial attacks against {ASR}.
\newblock \emph{IEEE Sign. Process. Letters}, 2024.

\bibitem[Baevski et~al.(2020)Baevski, Zhou, Mohamed, and Auli]{wav2vec}
Alexei Baevski, Yuhao Zhou, Abdelrahman Mohamed, and Michael Auli.
\newblock {wav2vec 2.0}: A framework for self-supervised learning of speech
  representations.
\newblock In \emph{NeurIPS}, 2020.

\bibitem[Radford et~al.(2023)]{whisper}
Alec Radford et~al.
\newblock Robust speech recognition via large-scale weak supervision.
\newblock In \emph{ICML}, 2023.

\bibitem[Abdullah et~al.(2022)]{sok}
Hadi Abdullah et~al.
\newblock {SoK}: The faults in our {ASRs}: An overview of attacks against
  automatic speech recognition and speaker identification systems.
\newblock In \emph{IEEE S\&P}, 2022.

\bibitem[Yakura and Sakuma(2019)]{robust}
Hiromu Yakura and Jun Sakuma.
\newblock Robust audio adversarial example for a physical attack.
\newblock In \emph{IJCAI}, 2019.

\bibitem[Wang et~al.(2023{\natexlab{a}})Wang, Zheng, Chen, Cheng, and
  Chen]{campp}
Hui Wang, Siqi Zheng, Yafeng Chen, Luyao Cheng, and Qian Chen.
\newblock {CAM++}: A fast and efficient network for speaker verification using
  context-aware masking.
\newblock In \emph{Interspeech}, 2023{\natexlab{a}}.

\bibitem[Chen et~al.(2023)Chen, Zheng, Wang, Cheng, Chen, and Qi]{eres2net}
Yafeng Chen, Siqi Zheng, Hui Wang, Luyao Cheng, Qian Chen, and Jiajun Qi.
\newblock An enhanced {Res2Net} with local and global feature fusion for
  speaker verification.
\newblock In \emph{Interspeech}, 2023.

\bibitem[Madry et~al.(2018)Madry, Makelov, Schmidt, Tsipras, and Vladu]{pgd}
Aleksander Madry, Aleksandar Makelov, Ludwig Schmidt, Dimitris Tsipras, and
  Adrian Vladu.
\newblock Towards deep learning models resistant to adversarial attacks.
\newblock In \emph{ICLR}, 2018.

\bibitem[Wang et~al.(2021{\natexlab{b}})Wang, Mallya, and Liu]{talkinghead1kh}
Ting-Chun Wang, Arun Mallya, and Ming-Yu Liu.
\newblock One-shot free-view neural talking-head synthesis for video
  conferencing.
\newblock In \emph{CVPR}, 2021{\natexlab{b}}.

\bibitem[Zhu et~al.(2022)Zhu, Wu, Zhu, Jiang, Tang, Zhang, Liu, and
  Loy]{celebvhq}
Hao Zhu, Wayne Wu, Wentao Zhu, Liming Jiang, Siwei Tang, Li~Zhang, Ziwei Liu,
  and Chen~Change Loy.
\newblock {CelebV-HQ}: A large-scale video facial attributes dataset.
\newblock In \emph{ECCV}, 2022.

\bibitem[Wang et~al.(2023{\natexlab{b}})Wang, Chan, and Loy]{clipiqa}
Jianyi Wang, Kelvin~C.K. Chan, and Chen~Change Loy.
\newblock Exploring {CLIP} for assessing the look and feel of images.
\newblock In \emph{AAAI}, 2023{\natexlab{b}}.

\bibitem[Taal et~al.(2010)Taal, Hendriks, Heusdens, and Jensen]{stoi}
Cees~H. Taal, Richard~C. Hendriks, Richard Heusdens, and Jesper Jensen.
\newblock A short-time objective intelligibility measure for time-frequency
  weighted noisy speech.
\newblock In \emph{ICASSP}, 2010.

\bibitem[Rix et~al.(2001)Rix, Beerends, Hollier, and Hekstra]{pesq}
Antony~W. Rix, John~G. Beerends, Michael~P. Hollier, and Andries~P. Hekstra.
\newblock Perceptual evaluation of speech quality ({PESQ}): An objective method
  for end-to-end speech quality assessment of narrow-band telephone networks
  and speech codecs.
\newblock In \emph{ICASSP}, 2001.

\bibitem[Chung and Zisserman(2016)]{syncnet}
Joon~Son Chung and Andrew Zisserman.
\newblock Out of time: Automated lip sync in the wild.
\newblock In \emph{ACCV}, 2016.

\bibitem[Chen et~al.(2019)Chen, Maddox, Duan, and Xu]{atvgnet}
Lele Chen, Ross~K. Maddox, Zhiyao Duan, and Chenliang Xu.
\newblock Hierarchical cross-modal talking face generation with dynamic
  pixel-wise loss.
\newblock In \emph{CVPR}, 2019.

\bibitem[Heusel et~al.(2017)Heusel, Ramsauer, Unterthiner, Nessler, and
  Hochreiter]{fid}
Martin Heusel, Hubert Ramsauer, Thomas Unterthiner, Bernhard Nessler, and Sepp
  Hochreiter.
\newblock {GANs} trained by a two time-scale update rule converge to a local
  nash equilibrium.
\newblock In \emph{NeurIPS}, 2017.

\bibitem[Unterthiner et~al.(2018)Unterthiner, van Steenkiste, Kurach, Marinier,
  Michalski, and Gelly]{fvd}
Thomas Unterthiner, Sjoerd van Steenkiste, Karol Kurach, Raphael Marinier,
  Marcin Michalski, and Sylvain Gelly.
\newblock Towards accurate generative models of video: A new metric and
  evaluations.
\newblock \emph{arXiv preprint arXiv:1812.01717}, 2018.

\bibitem[Hussain et~al.(2021)Hussain, Neekhara, Dubnov, McAuley, and
  Koushanfar]{waveguard}
Shehzeen Hussain, Paarth Neekhara, Shlomo Dubnov, Julian McAuley, and Farinaz
  Koushanfar.
\newblock {WaveGuard}: Understanding and mitigating audio adversarial examples.
\newblock In \emph{USENIX Security 21}, 2021.

\bibitem[Rouard et~al.(2023)Rouard, Massa, and D{\'e}fossez]{demucs}
Simon Rouard, Francisco Massa, and Alexandre D{\'e}fossez.
\newblock Hybrid transformers for music source separation.
\newblock In \emph{ICASSP}, 2023.

\bibitem[Guo et~al.(2024)Guo, Wang, Chen, Wang, Zhang, Chen, Yan, and
  Xiao]{wavepurifier}
Hanqing Guo, Guangjing Wang, Bocheng Chen, Yuanda Wang, Xiao Zhang, Xun Chen,
  Qiben Yan, and Li~Xiao.
\newblock {WavePurifier}: Purifying audio adversarial examples via hierarchical
  diffusion models.
\newblock In \emph{ACM MobiCom}, 2024.

\bibitem[Tram{\`e}r et~al.(2020)Tram{\`e}r, Carlini, Brendel, and
  Madry]{tramer}
Florian Tram{\`e}r, Nicholas Carlini, Wieland Brendel, and Aleksander Madry.
\newblock On adaptive attacks to adversarial example defenses.
\newblock In H.~Larochelle, M.~Ranzato, R.~Hadsell, M.F. Balcan, and H.~Lin,
  editors, \emph{NeurIPS}, volume~33, pages 1633--1645. Curran Associates,
  Inc., 2020.

\bibitem[Athalye et~al.(2018)Athalye, Carlini, and Wagner]{bpda}
Anish Athalye, Nicholas Carlini, and David Wagner.
\newblock Obfuscated gradients give a false sense of security: Circumventing
  defenses to adversarial examples.
\newblock In \emph{ICML}, 2018.

\bibitem[Du et~al.(2024)Du, Shen, Zhang, et~al.]{cosyvoice}
Zhihao Du, Qian~Chen Shen, Kai Zhang, et~al.
\newblock {CosyVoice}: A scalable multilingual zero-shot text-to-speech
  synthesizer based on supervised semantic tokens.
\newblock \emph{arXiv preprint arXiv:2407.05407}, 2024.

\bibitem[Goren et~al.(2023)]{xtts}
Eren Goren et~al.
\newblock {XTTS}: A massively multilingual zero-shot text-to-speech model.
\newblock \url{https://github.com/coqui-ai/TTS}, 2023.

\bibitem[{Resemble AI}(2025)]{chatterbox}
{Resemble AI}.
\newblock {Chatterbox-TTS}.
\newblock \url{https://github.com/resemble-ai/chatterbox}, 2025.
\newblock GitHub repository.

\bibitem[Desplanques et~al.(2020)Desplanques, Thienpondt, and
  Demuynck]{ecapatdnn}
Brecht Desplanques, Jenthe Thienpondt, and Kris Demuynck.
\newblock {ECAPA-TDNN}: Emphasized channel attention, propagation and
  aggregation in {TDNN} based speaker verification.
\newblock In \emph{Interspeech}, 2020.

\end{thebibliography}
\bibliographystyle{unsrtnat}

\appendix
\renewcommand{\theHfigure}{\thefigure}
\renewcommand{\theHtable}{\thetable}
\renewcommand{\theHequation}{\theequation}

\newpage
 \setcounter{equation}{0}
 \renewcommand{\theequation}{A.\arabic{equation}}
 \setcounter{figure}{0}
 \renewcommand{\thefigure}{A.\arabic{figure}}
 \setcounter{table}{0}
 \renewcommand{\thetable}{A.\arabic{table}}

\section{Discussions and Limitations}

\subsection{Limitations and Future Work}
\label{sec:app_A1}
\noindent\textbf{Adaptive Adversaries.}
Although MUTE is designed to be robust against realistic transformations such as denoising and adaptive attacks, adversarial attack and defense inherently form an arms race. As more advanced protection methods against talking-head exploitation are developed, new adaptive attacks may emerge to bypass them, potentially weakening the effectiveness of our approach. Future work will focus on developing more robust data protection strategies that can adapt to and defend against such emerging adaptive attacks.

\noindent\textbf{Audio-Only Limitation and Multimodal Extension.}
MUTE focuses on audio-only perturbations. Although combining MUTE with image-based methods such as Silencer~\cite{silencer} and MIS~\cite{syncbreaker} empirically improves performance, the two approaches are not jointly optimized and do not share a unified objective. In particular, existing methods differ in how failure modes are defined, such as suppressing motion versus inducing incorrect motion. Future work will explore unified multimodal optimization frameworks that jointly consider audio and visual perturbations under a consistent objective, potentially leading to stronger protection.

\noindent\textbf{Perceptual Quality Trade-off.}
While MUTE aims to preserve perceptual audio quality, there exists an inherent trade-off between perceptual fidelity and attack effectiveness. Stronger perturbations can improve alignment disruption but may degrade perceptual quality. Future work will explore perceptually-aware optimization techniques that better balance this trade-off, aiming to maintain high perceptual quality without sacrificing protection performance.

\noindent\textbf{Limited Evaluation Scope.}
The experiments are conducted on a limited number of samples per dataset, following prior work due to the high computational cost of talking-head generation. As a result, the evaluation may not fully capture the diversity of real-world scenarios. Future work will extend the evaluation to larger datasets and more diverse conditions to better assess generalization.

\noindent\textbf{Real-World Deployment Constraints.}
Although MUTE is designed with realistic threat models in mind, including black-box settings and robustness to transformations, practical deployment in real-world scenarios remains challenging. The method requires modifying audio prior to potential misuse and cannot protect already leaked audio. In addition, many real-world voice leakage scenarios involve real-time communication, such as phone calls. Future work will investigate efficient perturbation generation methods for real-time or low-latency settings to improve practical applicability.

\subsection{Subjective Bias}
\label{sec:app_A2}
The user study relies on subjective human perception, which may introduce inherent bias in the evaluation. In particular, the task is designed to identify the most noticeable lip-sync mismatch, rather than directly measuring the ability to prevent deepfake misuse. As a result, the evaluation focuses on perceptual misalignment rather than the broader objective of misuse prevention. In addition, individual differences in human perception and sensitivity to audio-visual mismatch may lead to variability in responses. The study also uses a limited number of samples, which may restrict coverage of diverse scenarios. To mitigate these effects, we report user study results with 95\% confidence intervals computed over all responses (N=1000), and assess statistical significance using a binomial test against random selection.

\subsection{Broader impacts}
\label{sec:app_A3}
This work aims to develop a proactive defense technique to protect individuals from voice misuse in talking-head generation systems. By disrupting audio-visual alignment, our approach contributes to mitigating potential deepfake exploitation and mitigating potential deepfake exploitation and strengthening proactive protection for multimodal data. We plan to release our method as an open-source implementation, which may serve as a useful baseline for future research in audio-based protection and adversarial robustness. However, the proposed method may also introduce potential risks. As an adversarial perturbation technique, it could be misused to intentionally degrade or manipulate audio-visual content beyond defensive purposes. In addition, similar techniques could be extended to develop more advanced attacks, contributing to the ongoing arms race between attack and defense methods.

 \setcounter{equation}{0}
 \renewcommand{\theequation}{B.\arabic{equation}}
 \setcounter{figure}{0}
 \renewcommand{\thefigure}{B.\arabic{figure}}
 \setcounter{table}{0}
 \renewcommand{\thetable}{B.\arabic{table}}
 
\section{Algorithm Details}
\label{sec:app_B}
\begin{algorithm}[h]
\caption{MUTE Adversarial Audio Generation}
\label{alg:mute}
\begin{algorithmic}[1]
\Require Clean audio $x$, reference image $I^{ref}$, audio encoder $f_{enc}$, perturbation budget $\epsilon$, step size $\alpha$,
PGD iterations $K$, phase ratio $\rho$, STFT loss weight $\lambda$
\Ensure Protected audio $x^{adv}$

\State Initialize $\delta_0 \sim \mathcal{U}(-\epsilon,\epsilon)$
\State $x^{adv}_0 \leftarrow \operatorname{clip}(x+\delta_0, x-\epsilon, x+\epsilon)$
\State Extract clean temporal embeddings
$z=f_{enc}(x)$
\State Obtain the target attention distribution $A^{target}$
\State Construct the spectral mask $M$ from $x$

\For{$k=0$ to $K-1$}
    \State $\delta_k \leftarrow x^{adv}_k-x$
    \State $\mathcal{L}_{\mathrm{stft}}
    \leftarrow
    \|\delta_k-\operatorname{ISTFT}(M\odot\operatorname{STFT}(\delta_k))\|_2^2$

    \If{$k \le \rho K$}
        \State Obtain adversarial attention distribution $A^{adv}$ from cross-attention
        \State $\mathcal{L}_{attn}
        \leftarrow
        \operatorname{KL}(A^{target}\|A^{adv})$
        \State $\mathcal{L}_{total}^{(k)}
        \leftarrow
        \mathcal{L}_{attn}
        +\lambda\mathcal{L}_{\mathrm{stft}}$
        
    \Else
        \State Extract adversarial temporal embeddings
        $z^{adv}=f_{enc}(x^{adv}_k)$
        \State $\mathcal{L}_{emb}
        \leftarrow
        \frac{1}{T}\sum_{t=1}^{T}
        \cos(z_t^{adv}, z_t)$
        \State $\mathcal{L}_{total}^{(k)}
        \leftarrow
        \mathcal{L}_{emb}
        +\lambda\mathcal{L}_{\mathrm{stft}}$
    \EndIf

    \State $g_k \leftarrow \nabla_{x^{adv}_k}\mathcal{L}_{total}^{(k)}$
    \State $x^{adv}_{k+1}
    \leftarrow
    \operatorname{clip}
    \left(
    x^{adv}_k-\alpha\cdot\operatorname{sign}(g_k),
    x-\epsilon,
    x+\epsilon
    \right)$
\EndFor

\State \Return $x^{adv}_K$
\end{algorithmic}
\end{algorithm}

Algorithm~\ref{alg:mute} summarizes the full optimization procedure of MUTE. We begin by initializing a perturbation $\delta_0$ within the $\epsilon$-bounded constraint and projecting the perturbed audio $x^{adv}_0$ into the valid range. The clean temporal embeddings $z = f_{enc}(x)$ are then extracted, and the target attention distribution $A^{target}$ along with the spectral mask $M$ are precomputed.

The adversarial audio is optimized through an iterative PGD procedure over $K$ steps. At each iteration $i$, we compute the current perturbation $\delta_i$ and apply the STFT loss, which constrains the perturbation to lie within a predefined spectral subspace. This encourages the perturbation to align with perceptually relevant regions of the speech signal, improving robustness against post-processing and denoising.

A key component of MUTE is the two-phase optimization strategy. During the early stage ($i < \rho K$), the objective focuses on \emph{Alignment-level Disruption}. Specifically, the adversarial attention map $A^{adv}$ is computed from cross-attention and optimized using the Target Attention KL Loss. This guides the attention distribution toward a predefined target, effectively disrupting audio-visual alignment and improving transferability across unseen models. In the later stage, the objective switches to \emph{Representation-level Degradation}, where adversarial embeddings $z^{adv}$ are extracted and optimized using a Frame-wise Embedding Cosine Loss, encouraging each temporal embedding to deviate from its clean counterpart.

At each iteration, the total loss is computed by combining the task-specific objective with the STFT loss. The adversarial audio is updated using the sign of the gradient, followed by projection to enforce the $\epsilon$-bounded constraint. After $K$ iterations, the final adversarial audio $x^{adv}_K$ is obtained.

Overall, MUTE unifies alignment-level disruption, representation-level degradation, and spectral-domain constraints within a single PGD-based framework, enabling effective and robust protection against talking-head generation.

 \setcounter{equation}{0}
 \renewcommand{\theequation}{C.\arabic{equation}}
 \setcounter{figure}{0}
 \renewcommand{\thefigure}{C.\arabic{figure}}
 \setcounter{table}{0}
 \renewcommand{\thetable}{C.\arabic{table}}
 
\section{Detailed Analysis}
\label{sec:app_C}

\subsection{Alignment-level Disruption}

\begin{table}[h]
\centering
\caption{Comparison of different sensitivity score formulations $s(n)$ on audio-visual alignment degradation across multiple models.}
\label{tab:sensitivity_score_comparison}
\renewcommand{\arraystretch}{1.8} 
\begin{tabular}{l cc | cc | cc}
\toprule
\multirow{2}{*}{$s(n)$} & \multicolumn{2}{c}{Hallo\cite{hallo}} & \multicolumn{2}{c}{Audio2Head\cite{audio2head}} & \multicolumn{2}{c}{SadTalker\cite{sadtalker}} \\
\cmidrule(lr){2-3} \cmidrule(lr){4-5} \cmidrule(lr){6-7}
& Sync$\downarrow$ & M-LMD$\uparrow$ & Sync$\downarrow$ & M-LMD$\uparrow$ & Sync$\downarrow$ & M-LMD$\uparrow$ \\
\midrule
$\displaystyle \frac{1}{n}$          & 1.485 & 1.511 & 3.963 & 1.217 & 5.083 & 0.756 \\
$\displaystyle \frac{1}{A(n)}$       & 1.407 & 1.571 & 3.922 & 1.231 & 5.043 & 0.761 \\
$\displaystyle \frac{|A'(n)|}{A(n)}$ & 1.383 & 1.608 & 3.813 & 1.255 & 4.957 & 0.795 \\
\rowcolor{gray!10} $\displaystyle \frac{|A'(n)| + \frac{\epsilon}{2}|A''(n)|}{A(n)}$ & \textbf{1.332} & \textbf{1.627} & \textbf{3.627} & \textbf{1.299} & \textbf{4.703} & \textbf{0.864} \\
\bottomrule
\end{tabular}
\end{table}

We construct the target attention distribution based on the sensitivity of each spatial location to small perturbations. To justify the proposed formulation of $s(n)$, we provide both theoretical motivation and empirical comparisons with alternative sensitivity scores.

Let $A(n)$ denote the attention value at spatial location $n$. Under an $\epsilon$-bounded perturbation, the variation of attention can be approximated via a Taylor expansion as:
\begin{equation}
A_\delta(n)
\approx
A(n) + \epsilon A'(n) + \frac{\epsilon^2}{2}A''(n),
\end{equation}
where $A'(n)$ and $A''(n)$ denote the first- and second-order sensitivities with respect to the perturbation. The resulting change in attention can be bounded as:
\begin{equation}
|A_\delta(n) - A(n)|
\le
\epsilon |A'(n)| + \frac{\epsilon^2}{2}|A''(n)|.
\end{equation}

This bound indicates that locations with larger $|A'(n)|$ or $|A''(n)|$ are more sensitive to perturbations and thus can induce larger changes in attention. Therefore, assigning higher importance to such locations is beneficial for effective alignment disruption.

However, considering only the absolute magnitude of change may overly favor locations with large attention values. To account for relative sensitivity, we normalize the change by the original attention value $A(n)$. Furthermore, since our goal is to compare relative importance across locations, we remove the global scaling factor by normalizing the bound with respect to $\epsilon$. This leads to the following sensitivity score:
\begin{equation}
s(n)
\propto
\frac{
|A'(n)| + \frac{\epsilon}{2}|A''(n)|
}{
A(n)
}.
\end{equation}

The target attention distribution is then defined by applying a softmax function:
\begin{equation}
A^{target}(n) = \operatorname{softmax}(s(n)).
\end{equation}

This formulation assigns higher probability mass to locations that are more sensitive to perturbations, enabling the optimization to focus on regions where the alignment structure can be more effectively disrupted. As a result, the attention map is redistributed toward structurally vulnerable regions, leading to stronger degradation of audio-visual alignment and improved transferability across different models.

To further validate the proposed formulation, we compare it with several alternative sensitivity scores, including a uniform weighting $\frac{1}{n}$, an inverse-attention weighting $\frac{1}{A(n)}$, and a first-order-only formulation $\frac{|A'(n)|}{A(n)}$. As shown in Table~\ref{tab:sensitivity_score_comparison}, our proposed $s(n)$ consistently achieves the best performance in both white-box and black-box settings. This demonstrates that incorporating both first- and second-order sensitivities provides a more effective and robust criterion for alignment disruption compared to simpler heuristics.

\subsection{Representation-level Degradation}

\begin{table}[h]
\centering
\caption{Comparison between global and frame-wise embedding cosine losses.}
\label{tab:global and frame-wise}
\begin{tabular}{lcc}
\toprule
 & Sync$\downarrow$ & M-LMD$\uparrow$ \\
\midrule
Global     & 2.484 & 1.317 \\
\rowcolor{gray!10} Frame-wise & \textbf{1.332} & \textbf{1.627} \\
\bottomrule
\end{tabular}
\end{table}

We analyze the effectiveness of the proposed frame-wise embedding cosine loss by comparing it with a global embedding cosine loss, both theoretically and empirically.

To illustrate the limitation of the global embedding cosine loss, consider the global representations:
\begin{equation}
\bar{z} = \frac{1}{T}\sum_{t=1}^{T} z_t, 
\qquad
\bar{z}^{adv} = \frac{1}{T}\sum_{t=1}^{T} z_t^{adv}.
\end{equation}
The corresponding global loss is defined as:
\begin{equation}
\mathcal{L}_{\text{global}} =
\cos \left( \bar{z}^{adv}, \bar{z} \right).
\end{equation}

The difference between the global embeddings can be written as:
\begin{equation}
\bar{z}^{adv} - \bar{z}
=
\frac{1}{T}
\sum_{t=1}^{T}
\left( z_t^{adv} - z_t \right),
\end{equation}
which leads to
\begin{equation}
\left\|
\bar{z}^{adv} - \bar{z}
\right\|_2
\le
\frac{1}{T}
\sum_{t=1}^{T}
\left\|
z_t^{adv} - z_t
\right\|_2.
\end{equation}

This inequality shows that the magnitude of the global embedding difference can be smaller than the average of per-frame differences due to averaging. As a result, perturbations at different time steps may partially cancel each other under the global embedding cosine loss, reducing the overall deviation.

In contrast, the proposed frame-wise embedding cosine loss directly penalizes each temporal component, preventing such averaging effects and enforcing consistent degradation across time.

As shown in Table~\ref{tab:global and frame-wise}, the proposed frame-wise embedding cosine loss leads to stronger degradation in audio-visual alignment compared to the global embedding cosine loss.

\subsection{STFT Loss Mask Design}

We provide additional details on the design of the STFT mask $M$ used in the loss formulation. Rather than applying perturbations uniformly across the spectrum, we focus on regions that are more likely to be preserved under real-world transformations.

Specifically, we restrict perturbations to the frequency range of 80–7600 Hz, corresponding to the primary speech band. This ensures that perturbations are applied within perceptually relevant frequency regions.

In addition, we apply a magnitude-based mask using a percentile threshold of 0.25 to focus on high-energy components of the signal. Since low-energy perturbations are more likely to be suppressed during filtering or denoising, this design improves the robustness of the perturbation.

Together, these design choices encourage perturbations to align with the underlying speech structure, improving robustness against post-processing and denoising.

\subsection{Optimization-aware Adaptive Attack Evaluations}
\label{supp:adaptive}
To assess the robustness of our protection against an adaptive adversary aware of
its design, we follow the adaptive-attack methodology of~\cite{tramer} and evaluate
it against an optimization-based purification attack that exploits the
speech band used by our protection, and handle its non-differentiable mask with
BPDA~\cite{bpda}. The same attack is applied identically to our method and all
baselines.

\noindent\textbf{Purification operator.}
Given a protected utterance $x_{\mathrm{adv}}$, the adversary optimizes an additive
correction $\eta$ ($\lVert\eta\rVert_\infty\le\epsilon$, $\epsilon=0.004$) and outputs
$\hat{x}=R(x_{\mathrm{adv}}+\eta)$. The operator $R$ computes
$S=\mathrm{STFT}(x_{\mathrm{adv}}+\eta)$ (1024-point FFT, hop 256) and selects
time--frequency bins with the binary mask
\[
M(t,f)=\mathbf{1}\!\left[|S(t,f)|\ge\tau\right]\cdot B(f),
\qquad \tau=\operatorname{Quantile}(|S|,0.25),
\]
where $\tau$ is computed over all time--frequency bins of $|S|$,
$B(f)=\mathbf{1}[80\,\mathrm{Hz}\le f\le7600\,\mathrm{Hz}]$ for methods employing a
speech-band mask, and $B(f)=1$ (full band) otherwise. The magnitudes of the selected
bins are replaced by a $3\times3$ time--frequency average while the phase of $S$ is
retained, and the result is inverted by the inverse STFT.

\noindent\textbf{Objective and optimization.}
We grant the adversary access to the clean utterance, which makes this an
oracle-assisted, attacker-favorable setting. The adversary minimizes
\[
\mathcal{L}(\eta)=-\frac{1}{T}\sum_{t=1}^{T}
\cos\!\big(E(\hat{x})_t,\,E(x_{\mathrm{clean}})_t\big),
\]
where $E$ denotes the wav2vec\,2.0 encoder conditioning the generator. The correction
$\eta$ is initialized to zero and updated by projected sign-gradient descent,
$\eta\leftarrow\mathrm{clip}_{[-\epsilon,\epsilon]}\big(\eta-\alpha\,\mathrm{sign}(\nabla_\eta\mathcal{L})\big)$,
with $\alpha=4\times10^{-4}$ for $K=80$ iterations.

\noindent\textbf{BPDA.}
The forward pass uses the hard mask $M$, while the backward pass replaces its
derivative with that of the sigmoid relaxation
\[
\widetilde{M}(t,f)=\sigma\!\left(\frac{|S(t,f)|-\tau}{\kappa}\right)\cdot B(f),
\qquad \kappa=0.1,
\]
in a straight-through manner. The band gate $B$ is fixed in both passes. Gradients
also reach $\eta$ through the unmasked bins and through the smoothed magnitude and
phase of the masked bins, which are differentiable; BPDA additionally provides a
surrogate gradient for the magnitude-based bin selection. Note that BPDA is applied only to the attacker's own masking operator $R$, not to any component of the protection, and the gradient through the generator's audio encoder $E$ is computed exactly.

\subsection{TTS-Based Resynthesis Evaluation}
\label{appendix:tts_texts}

We provide additional details on the TTS-based resynthesis experiments in Section~\ref{sec:tts_robustness}, including the evaluation setting, speaker verification protocol, and evaluation texts.

\paragraph{Evaluation Setting.}
We consider the case where an adversary uses the MUTE-protected clip itself as the speaker-reference input to a zero-shot TTS model and resynthesizes speech with arbitrary text. The attacker aims to preserve the original speaker characteristics while bypassing the direct use of the protected audio. We evaluate whether the optional speaker-level objective reduces speaker identity consistency under such transformations. This setting targets the protected clip itself and does not prevent cloning when the attacker possesses a separate unprotected recording of the same speaker.

\paragraph{Evaluation Metrics.}
We use the pretrained ECAPA-TDNN~\cite{ecapatdnn} model (\texttt{speechbrain/spkrec-ecapa-voxceleb}) as an independent speaker verification model. For each synthesized sample, we compute the cosine similarity between its speaker embedding and that of the corresponding original clean audio. Scores above 0.25 are classified as the same speaker, while scores at or below 0.25 are classified as different speakers and counted as defense successes. The DSR is defined as the proportion of synthesized samples classified as different speakers. The same verification model and fixed threshold are used for all TTS models and experimental conditions.

We additionally report Sim to capture continuous changes in speaker identity consistency that are not fully reflected by threshold-based DSR. These metrics evaluate whether TTS-based resynthesis preserves the original speaker characteristics. They do not directly measure human-perceived speaker similarity or establish that audio-visual misalignment persists after resynthesis.

\paragraph{Comparison Conditions.}
We compare unprotected audio, MUTE-protected audio without the speaker-level objective ($\lambda_{\text{spk}}=0$), and MUTE-protected audio with $\lambda_{\text{spk}}\in\{0.01,0.05,0.10\}$. The unprotected condition provides a reference for speaker identity consistency without protection, while the $\lambda_{\text{spk}}=0$ condition allows us to examine the contribution of the optional speaker-level objective. We also report SyncNet and M-LMD to evaluate its effect on the primary alignment-disruption performance. Lower SyncNet and higher M-LMD indicate stronger disruption of audio-visual alignment.

\paragraph{Evaluation Texts.}
For the TTS-based resynthesis experiments, we constructed a pool of 13 texts that simulate realistic voice-cloning misuse scenarios. From this pool, 5 texts were randomly sampled per protected audio clip to generate the 250 evaluation samples (50 audio clips $\times$ 5 texts). The same 250 reference--text pairs were used across all TTS models. The texts span three threat categories:

\begin{itemize}[leftmargin=*]
    \item \textbf{Financial fraud and impersonation} (3 texts):
    requests for money transfers based on fabricated emergencies or debts.
    \item \textbf{Account security phishing} (8 texts):
    fake security alerts, login verifications, and password reset requests.
    \item \textbf{Authority impersonation} (2 texts):
    fake emergency notices and identity verification demands.
\end{itemize}

\textbf{Ethical statement.} These texts are released solely for reproducibility and to characterize the realistic threat model that MUTE aims to mitigate. They are not intended to facilitate misuse.

The full list of texts is as follows:
\begin{enumerate}
    \item ``I urgently need money for an important surgery coming up soon, so please transfer me a thousand dollars.''
    \item ``I am in serious trouble while traveling in a foreign country, and I really need you to lend me some money.''
    \item ``You still owe me a thousand dollars from before, could you please send it to my new bank account?''
    \item ``There has been a severe emergency situation, and all residents are required to evacuate immediately.''
    \item ``Your recent bank transaction triggered a security alert. Please verify your identity to avoid unauthorized access.''
    \item ``We have detected multiple login attempts on your account. Please confirm your identity for security purposes.''
    \item ``Your credit card was used in a suspicious transaction. Can you confirm if this was authorized?''
    \item ``Your personal information might have been compromised. Reset your password to secure your account.''
    \item ``There is unusual activity on your account. Please verify your identity immediately to secure it.''
    \item ``A large withdrawal from your account has been flagged. Did you authorize this transaction?''
    \item ``To complete this transfer, we need you to confirm your social security number and banking details.''
    \item ``A suspicious login from an unknown device was detected. Please verify your identity to secure your records.''
    \item ``Your account has been temporarily locked after several failed logins. Please confirm your details to unlock it.''
\end{enumerate}

\subsection{Optimization}

\begin{table}[H]
\centering

\begin{minipage}{0.49\linewidth}
\centering
\caption{Effect of number of iterations $K$.}
\label{tab:iteration_effect}
\renewcommand{\arraystretch}{1.1}
\resizebox{\linewidth}{!}{
\begin{tabular}{lcc|ccc}
\toprule
$K$ & Sync$\downarrow$ & M-LMD$\uparrow$ & FID$\uparrow$ & FVD$\uparrow$ & E-FID$\uparrow$ \\
\midrule
50  & 1.817 & 1.539 & 2.808 & 1.970 & 5.031 \\
100 & 1.568 & 1.600 & 3.006 & 2.006 & 5.100 \\
150 & 1.458 & 1.623 & 3.167 & 2.254 & 5.124 \\
\rowcolor{gray!10} 200 & \textbf{1.332} & \textbf{1.627} & \textbf{3.183} & \textbf{2.363} & \textbf{5.489} \\
\bottomrule
\end{tabular}
}
\end{minipage}
\hfill
\begin{minipage}{0.49\linewidth}
\centering
\caption{Effect of step size $\alpha$.}
\vspace{-0.5mm}
\label{tab:performance_step_size}
\resizebox{\linewidth}{!}{
\begin{tabular}{lcc|ccc}
\toprule
$\alpha$ & Sync$\downarrow$ & M-LMD$\uparrow$ & FID$\uparrow$ & FVD$\uparrow$ & E-FID$\uparrow$ \\
\midrule
0.00005 & 2.024 & 1.564 & 2.481 & 1.822 & 4.783 \\
0.0001  & 1.613 & 1.577 & 2.931 & 2.030 & 5.253 \\
0.00015 & 1.455 & 1.590 & 2.881 & 2.053 & 5.318 \\
\rowcolor{gray!10} 
0.0002  & \textbf{1.332} & \textbf{1.627} & \underline{3.183} & \textbf{2.363} & \underline{5.489} \\
0.00025 & \underline{1.412} & \underline{1.623} & \textbf{3.234} & \underline{2.313} & \textbf{5.512} \\
\bottomrule
\end{tabular}
}
\end{minipage}

\end{table}

\noindent\textbf{Number of Iterations.}
To analyze the effect of the number of iterations, we conduct experiments by varying the total number of PGD steps among 50, 100, 150, and 200, while fixing the phase ratio to $\rho=0.1$. 

As shown in Table~\ref{tab:iteration_effect}, increasing the number of iterations consistently strengthens the attack performance. Both audio-visual alignment degradation and visual quality degradation improve as the number of iterations increases, as indicated by decreasing SyncNet scores and increasing M-LMD, FID, FVD, and E-FID values. These results suggest that additional optimization steps enable the model to more fully exploit the adversarial objective.

\noindent\textbf{Step Size.}
To analyze the effect of the step size, we conduct experiments by varying the step size from 0.00005 to 0.00025 while keeping other settings fixed. 

As shown in Table~\ref{tab:performance_step_size}, the alignment degradation improves as the step size increases from 0.00005 to 0.0002, achieving the best performance at 0.0002, and then degrades when further increased to 0.00025. A similar trend is observed in FVD. Although FID and E-FID reaches its best value at 0.00025, the overall performance is consistently strongest at 0.0002, indicating that a step size of 0.0002 provides the best balance between optimization effectiveness and stability.

\begin{table*}[h]
\centering
\caption{Effect of phase ratio $\rho$ and comparison with weighted-sum optimization.}
\label{tab:detailed_transfer}
\renewcommand{\arraystretch}{0.85}
\resizebox{\textwidth}{!}{%
\begin{tabular}{c| c | cc | c | cc | cc | cc}
\toprule
\multirow{2}{*}{$\rho$} & & Level 1 & Level 2 & \multirow{2}{*}{Time$\downarrow$} & \multicolumn{2}{c}{Hallo\cite{hallo}} & \multicolumn{2}{c}{Audio2Head\cite{audio2head}} & \multicolumn{2}{c}{SadTalker\cite{sadtalker}} \\
\cmidrule(lr){6-7} \cmidrule(lr){8-9} \cmidrule(lr){10-11}
& & Loss Red. & Loss Red. & & Sync$\downarrow$ & M-LMD$\uparrow$ & Sync$\downarrow$ & M-LMD$\uparrow$ & Sync$\downarrow$ & M-LMD$\uparrow$ \\
\midrule
\multirow{2}{*}{1.0} 
& $L_{emb}$ & \multicolumn{2}{c|}{-44.34\%} & \multirow{2}{*}{78.354} & \multirow{2}{*}{3.920} & \multirow{2}{*}{0.997} & \multirow{2}{*}{3.958} & \multirow{2}{*}{1.163} & \multirow{2}{*}{5.048} & \multirow{2}{*}{0.616} \\
& $L_{attn}$  & \multicolumn{2}{c|}{-96.43\%} & & & & & & & \\
\midrule
\multirow{2}{*}{0.2} 
& $L_{emb}$ & -95.13\% & -82.56\% & \multirow{2}{*}{20.087} & \multirow{2}{*}{\underline{1.419}} & \multirow{2}{*}{\underline{1.578}} & \multirow{2}{*}{\underline{3.798}} & \multirow{2}{*}{1.241} & \multirow{2}{*}{4.898} & \multirow{2}{*}{0.698} \\
& $L_{attn}$  & -36.63\% & -95.45\% & & & & & & & \\
\midrule
\rowcolor{gray!10} & $L_{emb}$ & -95.45\% & -84.34\% & & & & & & & \\
\rowcolor{gray!10} \multirow{-2}{*}{0.1} 
& $L_{attn}$  & -40.12\% & -91.87\% & \multirow{-2}{*}{12.345} & \multirow{-2}{*}{\textbf{1.332}} & \multirow{-2}{*}{\textbf{1.627}} & \multirow{-2}{*}{\textbf{3.627}} & \multirow{-2}{*}{\textbf{1.299}} & \multirow{-2}{*}{\textbf{4.703}} & \multirow{-2}{*}{\textbf{0.864}} \\
\midrule
\multirow{2}{*}{0.05} 
& $L_{emb}$ & -97.45\% & -89.74\% & \multirow{2}{*}{9.198} & \multirow{2}{*}{1.470} & \multirow{2}{*}{1.551} & \multirow{2}{*}{3.799} & \multirow{2}{*}{\underline{1.246}} & \multirow{2}{*}{4.899} & \multirow{2}{*}{0.702} \\
& $L_{attn}$  & -42.02\% & -88.72\% & & & & & & & \\
\midrule
\multirow{2}{*}{0.0} 
& $L_{emb}$ & \multicolumn{2}{c|}{-96.92\%} & \multirow{2}{*}{4.824} & \multirow{2}{*}{1.427} & \multirow{2}{*}{1.520} & \multirow{2}{*}{3.904} & \multirow{2}{*}{1.232} & \multirow{2}{*}{5.063} & \multirow{2}{*}{0.698} \\
& $L_{attn}$  & \multicolumn{2}{c|}{-43.88\%} & & & & & & & \\
\midrule
\midrule
\multirow{2}{*}{\shortstack{weighted\\sum}} 
& $L_{emb}$ & \multicolumn{2}{c|}{-89.30\%} & \multirow{2}{*}{79.548} & \multirow{2}{*}{1.512} & \multirow{2}{*}{1.532} & \multirow{2}{*}{3.911} & \multirow{2}{*}{1.190} & \multirow{2}{*}{4.940} & \multirow{2}{*}{0.715} \\
& $L_{attn}$  & \multicolumn{2}{c|}{-89.14\%} & & & & & & & \\
\bottomrule
\end{tabular}
}
\end{table*}

\noindent\textbf{Optimization Strategy Analysis}
Within the optimization strategy, we analyze how the phase ratio affects both loss reduction and alignment degradation performance. We also compare the proposed two-phase optimization with a weighted-sum approach.

As shown in Table~\ref{tab:detailed_transfer}, when optimizing only a single objective, both $\mathcal{L}_{attn}$ ($\rho=1.0$) and $\mathcal{L}_{emb}$ ($\rho=0.0$) can be reduced by over 96\%. However, these settings exhibit different behaviors: optimizing only $\mathcal{L}_{attn}$ results in weaker white-box performance but stronger transferability, while optimizing only $\mathcal{L}_{emb}$ shows the opposite trend. When using two-phase optimization, we observe that reducing one objective often increases the other, indicating conflicting optimization directions. This makes it important to find a proper balance between the two objectives, where a phase ratio of 1:9 achieves the best overall performance in our experiments. Although the weighted-sum method also achieves strong loss reduction and competitive performance, the two-phase strategy provides better efficiency when considering computational cost, making it more effective for the proposed method.

 \setcounter{equation}{0}
 \renewcommand{\theequation}{D.\arabic{equation}}
 \setcounter{figure}{0}
 \renewcommand{\thefigure}{D.\arabic{figure}}
 \setcounter{table}{0}
 \renewcommand{\thetable}{D.\arabic{table}}

 \begin{figure*}[t]
	\centering
  \includegraphics[width=1.0\linewidth]{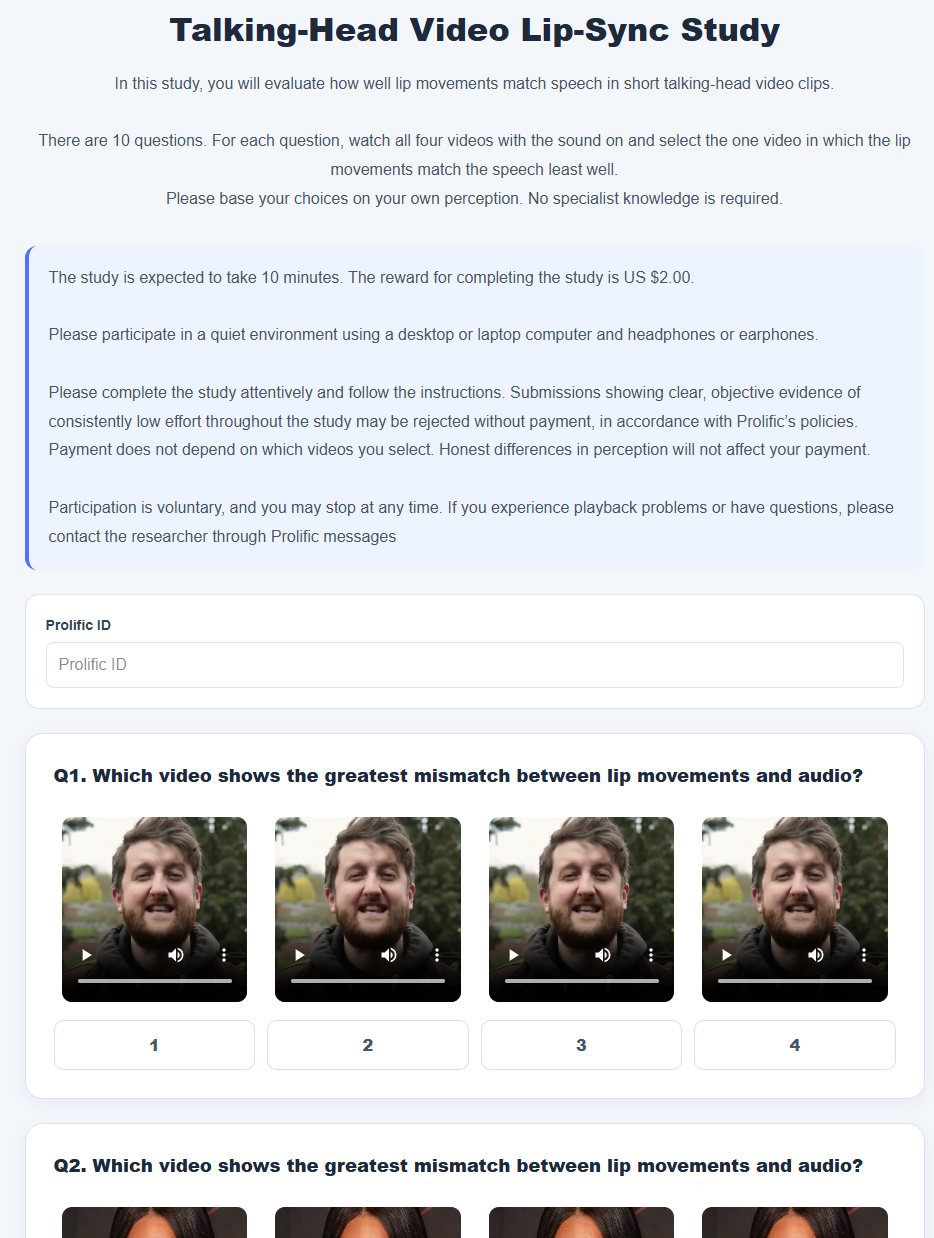}
  \caption{User study interface and instructions.}
  \label{fig:suppl_user_study}
  \vspace{-3mm}
\end{figure*}
 
\section{User Study}
\label{sec:app_D}
We conducted a user study to evaluate whether the degradation of lip synchronization introduced by MUTE is perceptually noticeable to human observers. The study was carried out using the Prolific~\footnote{\url{https://www.prolific.com}} platform for participant recruitment and compensation. No personally identifiable information was collected, and all participation was conducted anonymously through the platform. Participants were provided with clear instructions and asked to select, among multiple videos, the one in which the lip movements and audio appeared to be the most mismatched. The full instruction text and screenshots of the experimental interface are provided in Fig.~\ref{fig:suppl_user_study}.

We recruited 100 participants and provided compensation of \$2 per participant, corresponding to approximately a 10-minute study duration. Each participant evaluated 10 samples, resulting in a total of 1,000 responses. The evaluation samples were randomly selected from the TalkingHead-1KH and CelebV-HQ datasets and included videos generated by MUTE and baseline methods. Results are reported as the proportion of selections for each method, along with 95\% confidence intervals computed over all responses (N = 1,000) to assess statistical reliability.

 \setcounter{equation}{0}
 \renewcommand{\theequation}{E.\arabic{equation}}
 \setcounter{figure}{0}
 \renewcommand{\thefigure}{E.\arabic{figure}}
 \setcounter{table}{0}
 \renewcommand{\thetable}{E.\arabic{table}}

\section{Additional Visualizations}
\label{sec:app_E}
\subsection{Qualitative Comparison on Benchmark Datasets}
\begin{figure*}[h]
	\centering
  \includegraphics[width=0.95\linewidth]{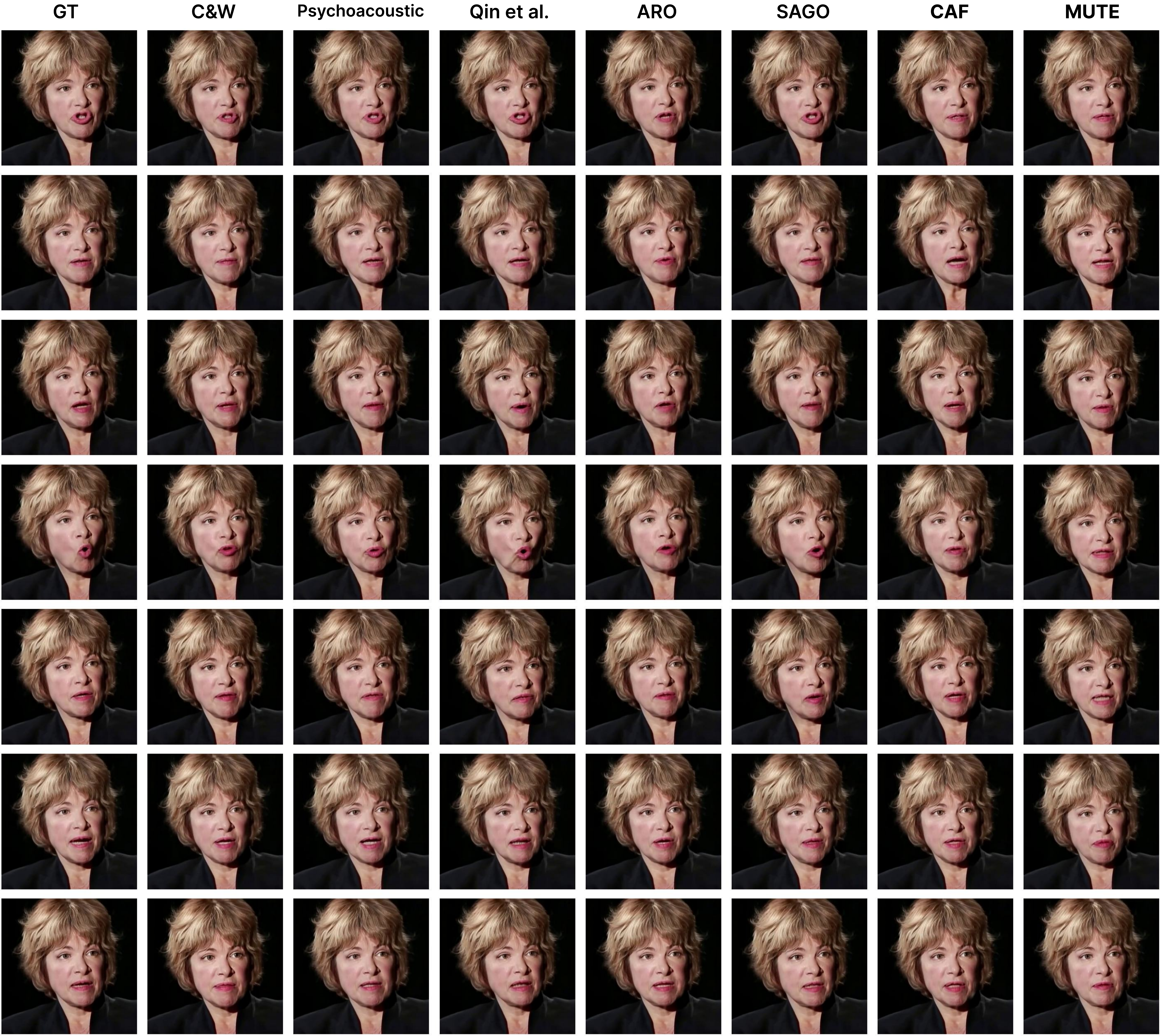}
  \vspace{-3mm}
  \caption{\textbf{Per-frame qualitative comparison on TalkingHead-1KH.} A frame-by-frame (row) visual comparison between baseline methods and MUTE, demonstrating MUTE's significant disruption of lip synchronization and facial dynamics.}
  \label{fig:qualitative1}
  \vspace{-3mm}
\end{figure*}

\begin{figure*}[h]
	\centering
  \includegraphics[width=0.95\linewidth]{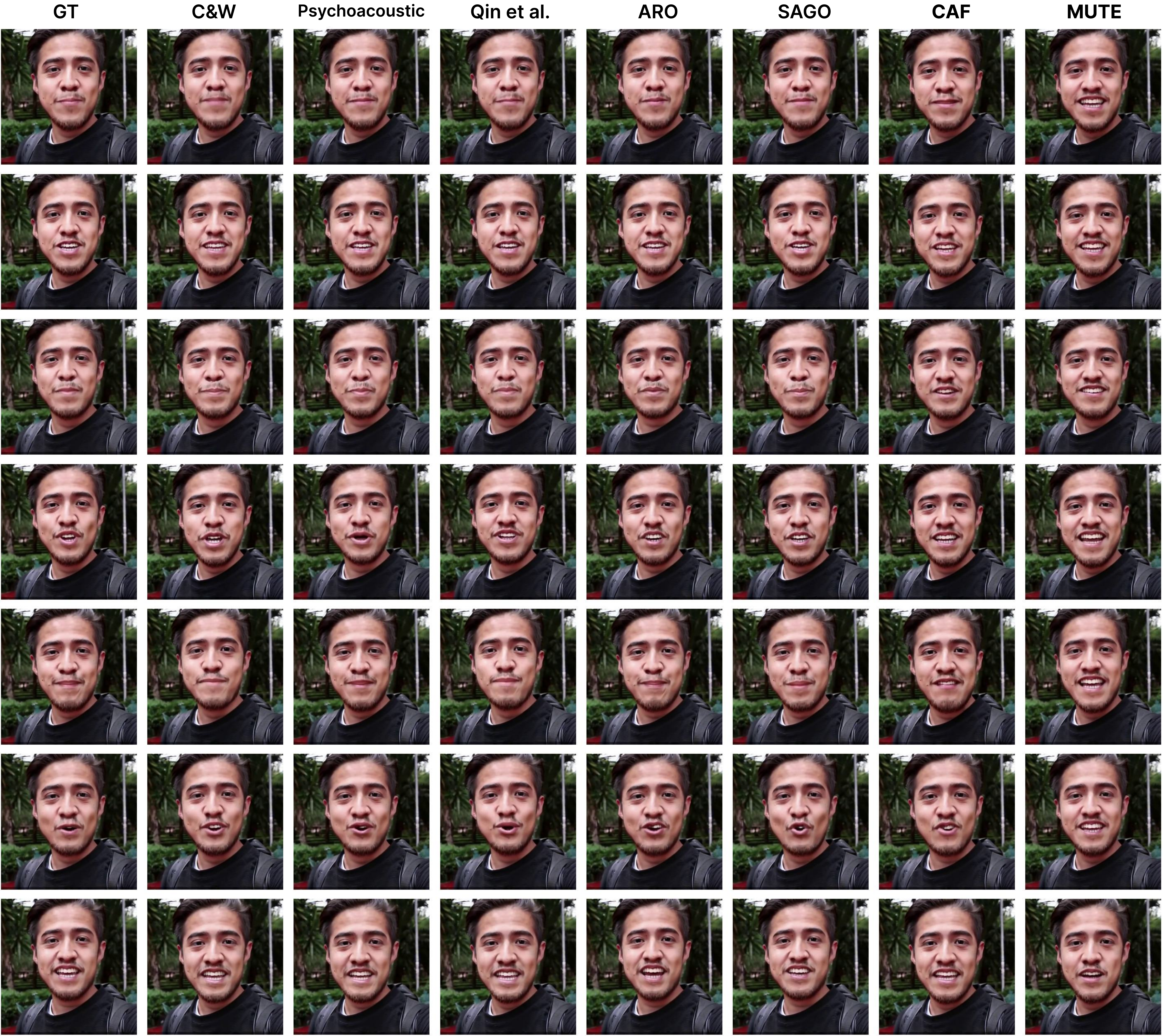}
  \vspace{-3mm}
  \caption{\textbf{Per-frame qualitative comparison on CelebV-HQ.} A frame-by-frame (row) visual comparison between baseline methods and MUTE, demonstrating MUTE's significant disruption of lip synchronization and facial dynamics.}
  \label{fig:qualitative2}
  \vspace{-3mm}
\end{figure*}

Fig.~\ref{fig:qualitative1} and Fig.~\ref{fig:qualitative2} present a qualitative comparison of MUTE against ground truth and various baseline methods on the TalkingHead-1KH and CelebV-HQ datasets, respectively. Compared to the original ground truth (GT) and existing adversarial attacks such as C\&W, Psychoacoustic, Qin et al., ARO, and SAGO, MUTE induces the most significant and perceptually noticeable disparity in lip synchronization. The facial animations generated with MUTE-protected audio exhibit lip shapes and motion intensities that are completely detached from the phonetic content of the driving speech. These visual results align with the quantitative superiority of MUTE in metrics like SyncNet and M-LMD.

\begin{figure*}[h]
	\centering
  \includegraphics[width=0.95\linewidth]{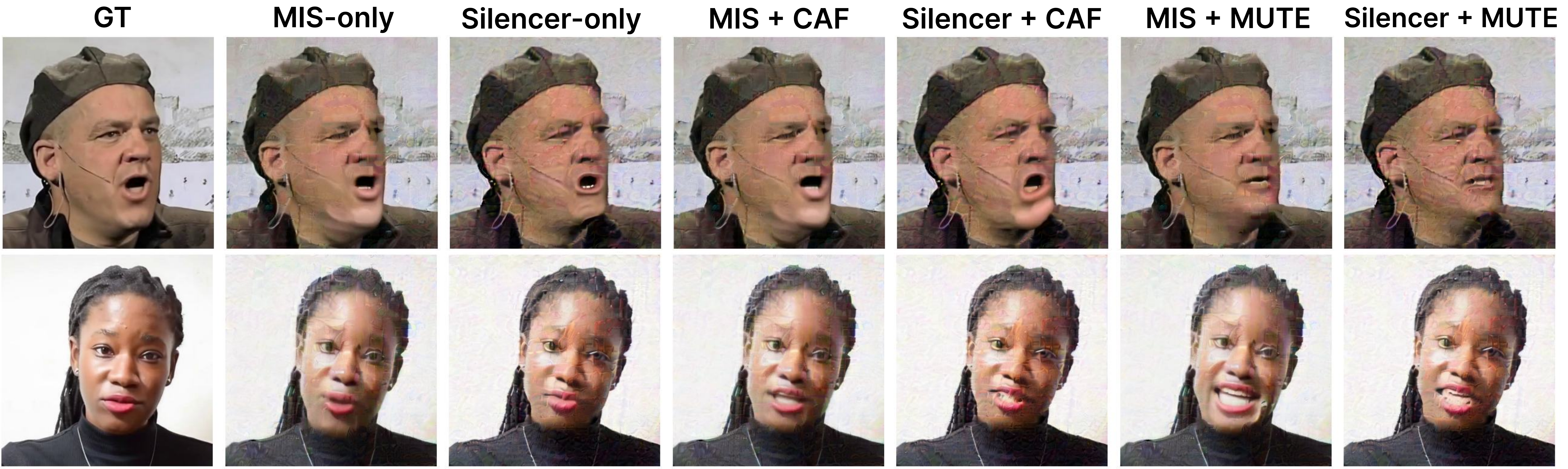}
  \vspace{-3mm}
    \caption{\textbf{Qualitative results with image-level protection.} MIS and Silencer are shown alone and combined with CAF or MUTE. Only the MUTE combinations produce mouth shapes that clearly deviate from the ground truth (GT).}
  \label{fig:qualitative3}
  \vspace{-3mm}
\end{figure*}

\subsection{Qualitative Results with Image-Level Protection}
In the main paper (Table~\ref{tab:silencer}), we showed that MUTE is complementary to image-level protection methods. Here we provide the corresponding qualitative comparison. Figure~\ref{fig:qualitative3} shows frames generated with MIS~\cite{syncbreaker} or Silencer~\cite{silencer} alone, and with each of them combined with CAF~\cite{syncbreaker} or MUTE.
Image-level protection alone introduces visual artifacts, but the mouth shape still follows the ground truth. Adding CAF changes this only slightly, and the lip motion remains largely aligned with the speech. In contrast, adding MUTE produces a clear mismatch in both directions: the mouth stays nearly closed where the ground truth is wide open (top), and opens where the ground truth is
closed (bottom). Meanwhile, the level of visual degradation is similar across all combinations, consistent with their comparable FID, FVD, and E-FID scores. These observations match the lower Sync and higher M-LMD scores of the MUTE combinations, and show that the two types of protection act on complementary aspects: image-level perturbations degrade visual quality, while MUTE disrupts audio-visual alignment.

\subsection{Performance on Black-Box Models}
\begin{figure*}[h]
	\centering
  \includegraphics[width=0.95\linewidth]{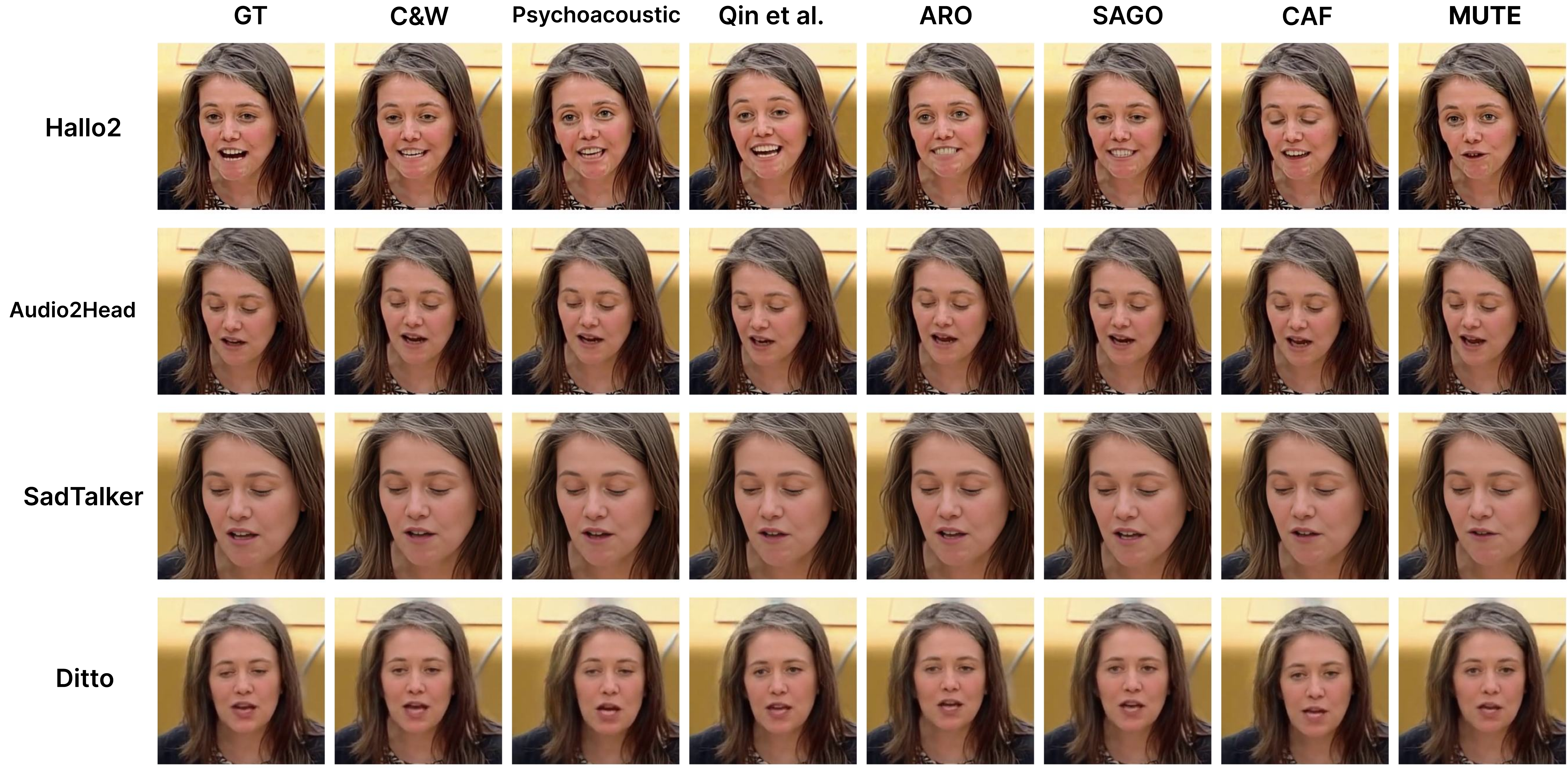}
  \vspace{-3mm}
  \caption{\textbf{Qualitative evaluation of transferability.} Visual comparison demonstrating MUTE's effective disruption of lip synchronization across unseen black-box talking-head architectures (rows: Hallo2, Audio2Head, SadTalker, Ditto).}
  \label{fig:qualitative4}
  \vspace{-3mm}
\end{figure*}

The transferability of MUTE to diverse talking-head architectures is visualized in Fig.~\ref{fig:qualitative4}. Despite the adversarial perturbations being generated using the surrogate Hallo model, they effectively disrupt the audio-visual alignment of various unseen target models, including Hallo2, Audio2Head, SadTalker, and Ditto. These results demonstrate that MUTE effectively targets model-agnostic alignment structures, ensuring strong protection even when the specific generation architecture used by an attacker is unknown. Unlike baselines that show inconsistent performance across different targets, MUTE maintains severe lip-sync degradation across all evaluated black-box scenarios.

\begin{figure*}[h]
	\centering
  \includegraphics[width=0.95\linewidth]{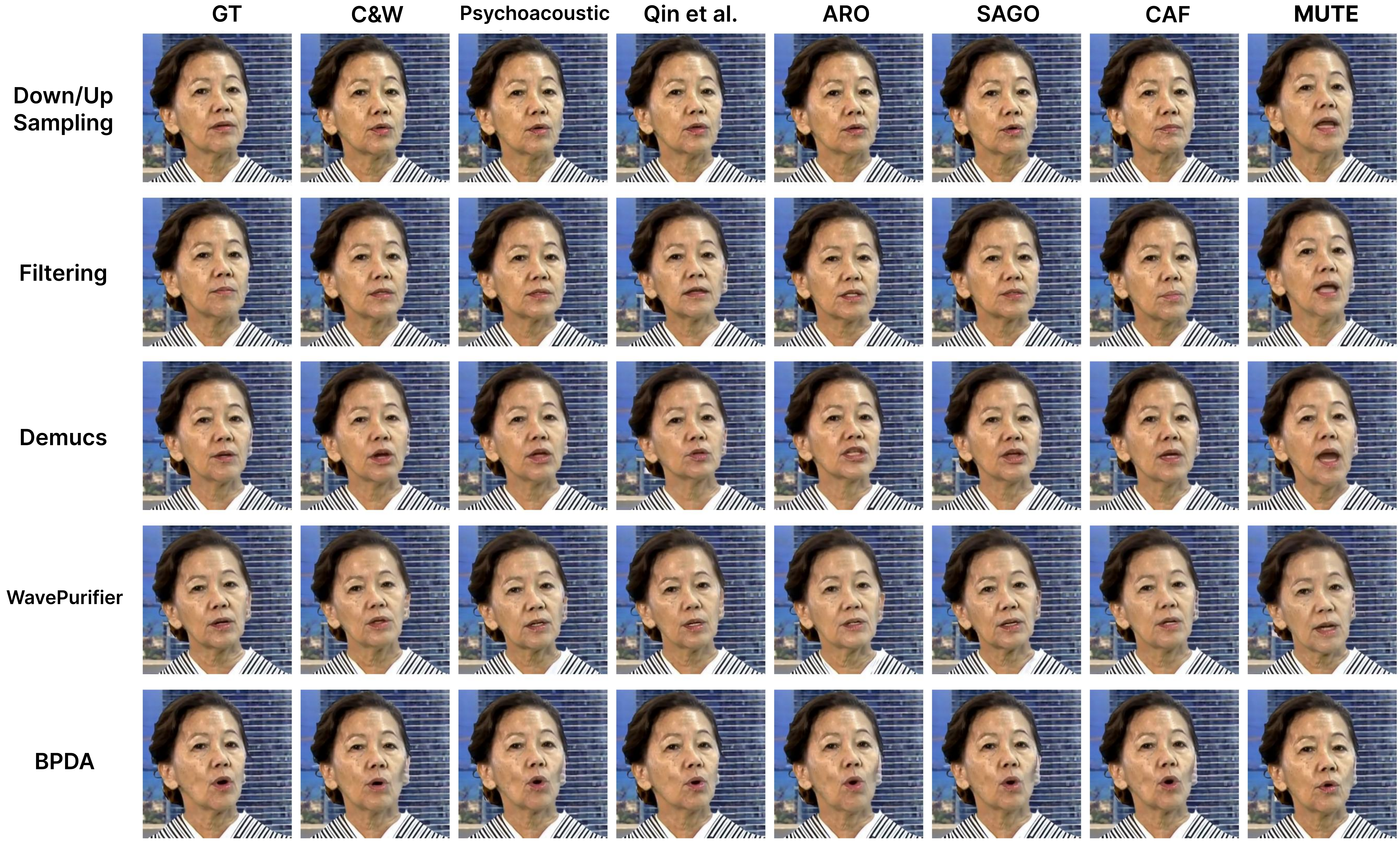}
  \vspace{-3mm}
  \caption{\textbf{Qualitative evaluation of robustness.} Visual comparison demonstrating MUTE's consistent disruption of lip synchronization even after audio is subjected to various real-world transformations and denoising attacks (rows: Down/Up Sampling, Filtering, Demucs, WavePurifier, Adaptive Attack).}
  \label{fig:qualitative5}
  \vspace{-3mm}
\end{figure*}

\subsection{Robustness to Post-processing, Denoising and Adaptive Attack}
Fig.~\ref{fig:qualitative5} illustrates the robustness of MUTE against common real-world transformations and model-based denoising techniques. The visualizations confirm that the protective effect of MUTE remains highly effective even after the audio has undergone operations such as sampling changes, spectral filtering, and advanced purification using Demucs or WavePurifier. This resilience is a direct result of our robustness-aware optimization in the STFT domain, which ensures that perturbations reside in high-energy regions that are difficult to remove without significantly degrading the quality of the underlying speech signal. These qualitative examples verify that MUTE provides a stable defense in realistic security environments where attackers may attempt to neutralize perturbations.

 \setcounter{equation}{0}
 \renewcommand{\theequation}{F.\arabic{equation}}
 \setcounter{figure}{0}
 \renewcommand{\thefigure}{F.\arabic{figure}}
 \setcounter{table}{0}
 \renewcommand{\thetable}{F.\arabic{table}}

\section{Additional Experiments}

\begin{table*}[h]
\centering
\caption{Purification performance when each purifier is restricted to the speech band (80--7600\,Hz) versus applied to the full band. Lower Sync and higher M-LMD indicate stronger retained protection.}
\label{tab:attack_band_setting}
\fontsize{7pt}{8.5pt}\selectfont
\renewcommand{\arraystretch}{0.95}
\setlength{\tabcolsep}{5pt}
\begin{tabular}{l cc | cc | cc | cc}
\toprule
\multirow{2}{*}{Purification Band}
  & \multicolumn{2}{c}{Down/Up~\cite{waveguard}}
  & \multicolumn{2}{c}{Filtering~\cite{waveguard}}
  & \multicolumn{2}{c}{Demucs~\cite{demucs}}
  & \multicolumn{2}{c}{WavePurifier~\cite{wavepurifier}} \\
\cmidrule(lr){2-3} \cmidrule(lr){4-5} \cmidrule(lr){6-7} \cmidrule(lr){8-9}
  & Sync$\downarrow$ & M-LMD$\uparrow$
  & Sync$\downarrow$ & M-LMD$\uparrow$
  & Sync$\downarrow$ & M-LMD$\uparrow$
  & Sync$\downarrow$ & M-LMD$\uparrow$ \\
\midrule
Speech-band (80--7600\,Hz)
  & \textbf{2.873} & \textbf{1.268} & 2.306 & \textbf{1.434} & \textbf{1.532} & 1.584 & \textbf{3.444} & \textbf{1.222} \\
Full-band
  & 2.914 & 1.254 & \textbf{2.251} & 1.410 & 1.534 & \textbf{1.593} & 3.494 & 1.217 \\
\bottomrule
\end{tabular}
\end{table*}

\subsection{Protection-Aware Speech-Band Attack}
We additionally evaluated a protection-aware attacker that specifically targets the 80--7600\,Hz range in which MUTE concentrates its perturbations. We restricted down/up-sampling, spectral filtering, Demucs, and WavePurifier to this band, thereby modeling an attacker who knows the frequency range used by MUTE's STFT mask. Table~\ref{tab:attack_band_setting} compares this setting with standard full-band purification.

As shown in Table~\ref{tab:attack_band_setting}, targeting the protected speech band does not make purification more effective. Across all four purifiers, the speech-band and full-band settings yield nearly identical Sync and M-LMD scores, with no consistent advantage for the speech-band attack. These results show that the protection is not concentrated in a narrow, trivially removable sub-band, and that knowing the frequency range used by the defense gives the attacker no meaningful advantage. Importantly, MUTE continued to outperform the evaluated baselines under all four attacks.

\begin{figure*}[h]
  \centering
  \includegraphics[width=0.95\linewidth]{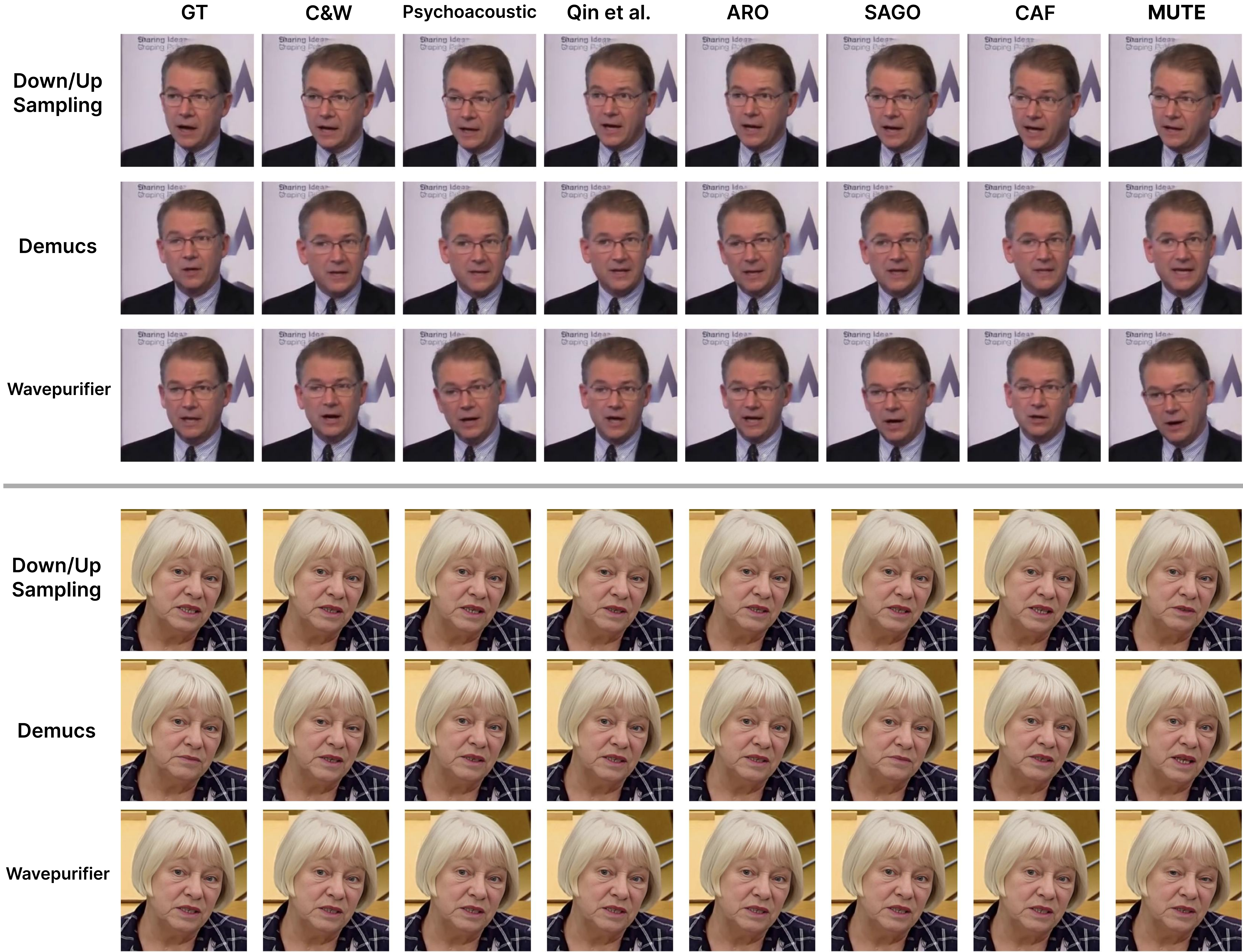}
  \vspace{-3mm}
  \caption{\textbf{Qualitative results under compound purification and black-box transfer.} The protected audio is purified and then used to drive Audio2Head (top) and SadTalker (bottom).}
  \label{fig:qualitative6}
  \vspace{-3mm}
\end{figure*}

\subsection{Compound Purification and Black-Box Evaluation}
Evaluating purification and black-box transfer independently does not fully capture the compound threat in which an attacker first purifies the protected audio and then applies it to an unseen talking-head model. Therefore, we additionally evaluate this setting by applying down/up-sampling, Demucs, or WavePurifier to the protected audio, followed by video generation using the black-box target models Audio2Head and SadTalker.

Table~\ref{tab:purification_transferability} reports the quantitative results. MUTE achieves the lowest Sync and the highest M-LMD for every combination of purifier and target model, consistently outperforming all baselines. This indicates that its protection survives purification and still transfers to unseen models. Figure~\ref{fig:qualitative6} shows qualitative examples for both target models, where the lip motion generated from purified MUTE-protected audio remains visibly disrupted.

\begin{table*}[t]
\centering
\caption{Transferability evaluation under various audio purification methods across talking-head models.}
\label{tab:purification_transferability}
\fontsize{7.5pt}{9.5pt}\selectfont
\renewcommand{\arraystretch}{0.95}
\setlength{\tabcolsep}{16pt} %

\begin{tabular}{ll cc | cc}
\toprule
\multirow{2}{*}{\begin{tabular}[c]{@{}l@{}}Purification\\ Method\end{tabular}} 
  & \multirow{2}{*}{Method} 
  & \multicolumn{2}{c}{Audio2Head~\cite{audio2head}} 
  & \multicolumn{2}{c}{SadTalker~\cite{sadtalker}} \\
\cmidrule(lr){3-4} \cmidrule(lr){5-6}
  & & Sync$\downarrow$ & M-LMD$\uparrow$ & Sync$\downarrow$ & M-LMD$\uparrow$ \\
\midrule
\multirow{7}{*}{Down/Up~\cite{waveguard}}
  & C\&W~\cite{carlini}          & 3.409 & 1.384 & 4.572 & 0.670 \\
  & Qin et al.~\cite{qin}       & 3.645 & 1.388 & 4.783 & 0.691 \\
  & Psychoacoustic~\cite{psychoacoustic} & 3.450 & 1.382 & 4.592 & 0.672 \\
  & ARO~\cite{aro}                & 3.514 & 1.337 & 4.581 & 0.716 \\
  & SAGO~\cite{sago}              & 3.580 & 1.297 & 4.628 & 0.613 \\
  & CAF~\cite{syncbreaker}        & 3.330 & 1.376 & 4.605 & 0.686 \\
  & \textbf{MUTE}                 & \textbf{3.204} & \textbf{1.414} & \textbf{4.335} & \textbf{0.756} \\
\midrule
\multirow{7}{*}{Demucs~\cite{demucs}}
  & C\&W~\cite{carlini}          & 4.386 & 1.184 & 5.549 & 0.612 \\
  & Qin et al.~\cite{qin}       & 4.535 & 1.184 & 5.608 & 0.597 \\
  & Psychoacoustic~\cite{psychoacoustic} & 4.383 & 1.148 & 5.510 & 0.601 \\
  & ARO~\cite{aro}                & 4.168 & 1.232 & 5.319 & 0.632 \\
  & SAGO~\cite{sago}              & 4.382 & 1.178 & 5.421 & 0.549 \\
  & CAF~\cite{syncbreaker}        & 4.349 & 1.183 & 5.556 & 0.632 \\
  & \textbf{MUTE}                 & \textbf{2.998} & \textbf{1.419} & \textbf{5.177} & \textbf{0.691} \\
\midrule
\multirow{7}{*}{WavePurifier~\cite{wavepurifier}}
  & C\&W~\cite{carlini}          & 5.042 & 1.320 & 4.360 & 0.856 \\
  & Qin et al.~\cite{qin}       & 5.026 & 1.285 & 4.340 & 0.852 \\
  & Psychoacoustic~\cite{psychoacoustic} & 4.986 & 1.318 & 4.348 & 0.859 \\
  & ARO~\cite{aro}                & 4.982 & 1.331 & 4.333 & 0.861 \\
  & SAGO~\cite{sago}              & 4.965 & 1.349 & 4.283 & 0.848 \\
  & CAF~\cite{syncbreaker}        & 4.830 & 1.365 & 4.331 & 0.862 \\
  & \textbf{MUTE}                 & \textbf{4.335} & \textbf{1.390} & \textbf{3.714} & \textbf{0.973} \\
\bottomrule
\end{tabular}
\end{table*}

\newpage
\clearpage
\section*{NeurIPS Paper Checklist}

\begin{enumerate}

\item {\bf Claims}
    \item[] Question: Do the main claims made in the abstract and introduction accurately reflect the paper's contributions and scope?
    \item[] Answer: \answerYes{}
    \item[] Justification: The main contributions are clearly presented in the Section~\ref{sec:method} and Section~\ref{sec:experiments}, and are consistent with the claims made in the abstract and introduction.

\item {\bf Limitations}
    \item[] Question: Does the paper discuss the limitations of the work performed by the authors?
    \item[] Answer: \answerYes{}
    \item[] Justification: The limitations of our work are discussed in Appendix~\ref{sec:app_A1}, where we outline key limitations such as adaptive adversaries, perceptual trade-offs, and deployment constraints, along with corresponding future research directions.

\item {\bf Theory assumptions and proofs}
    \item[] Question: For each theoretical result, does the paper provide the full set of assumptions and a complete (and correct) proof?
    \item[] Answer: \answerNA{}
    \item[] Justification: The paper does not include formal theoretical results such as theorems or proofs.

    \item {\bf Experimental result reproducibility}
    \item[] Question: Does the paper fully disclose all the information needed to reproduce the main experimental results of the paper to the extent that it affects the main claims and/or conclusions of the paper (regardless of whether the code and data are provided or not)?
    \item[] Answer: \answerYes{}
    \item[] Justification: We provide detailed descriptions of the model architecture, loss functions, and optimization procedure in the Section~\ref{sec:method}, and experimental settings including datasets, baselines, and hyperparameters in the Section~\ref{sec:experiments}. Additional implementation details, full algorithmic procedures, and extended analyses are provided in Appendix~\ref{sec:app_B} and Appendix~\ref{sec:app_C}, enabling reproduction of the main experimental results.

\item {\bf Open access to data and code}
    \item[] Question: Does the paper provide open access to the data and code, with sufficient instructions to faithfully reproduce the main experimental results, as described in supplemental material?
    \item[] Answer: \answerYes{} %
    \item[] Justification: Our source code is publicly available at
        \url{https://github.com/jindl465/MUTE}, along with a README describing the environment setup, execution instructions, and an example sample for quick verification. For data, we use publicly available datasets (TalkingHead-1KH and CelebV-HQ) following the protocols described in Section~\ref{sec:experiments}, which provide sufficient information to reproduce our experimental results.

\item {\bf Experimental setting/details}
    \item[] Question: Does the paper specify all the training and test details (e.g., data splits, hyperparameters, how they were chosen, type of optimizer) necessary to understand the results?
    \item[] Answer: \answerYes{}
    \item[] Justification: We provide detailed experimental settings including datasets, baselines, hyperparameters, and evaluation metrics in the Section~\ref{sec:experiments}. Additional implementation details are provided in the Appendix~\ref{sec:app_C}, ensuring that the experimental setup is clearly described and the results are interpretable.

\item {\bf Experiment statistical significance}
    \item[] Question: Does the paper report error bars suitably and correctly defined or other appropriate information about the statistical significance of the experiments?
    \item[] Answer: \answerYes{}
    \item[] Justification: The main experiments are repeated with three independent random seeds, and the results are reported as mean $\pm$ standard deviation. In addition, we report statistical significance for the user study using 95\% confidence intervals computed over all responses (N=1000).

\item {\bf Experiments compute resources}
    \item[] Question: For each experiment, does the paper provide sufficient information on the computer resources (type of compute workers, memory, time of execution) needed to reproduce the experiments?
    \item[] Answer: \answerYes{}
    \item[] Justification: We report the type of compute resource (NVIDIA RTX 4090 GPU) used for experiments. In addition, we provide relative runtime comparisons across different configurations in the ablation study (Table~\ref{tab:kl_emb_ratio}), which reflect the computational cost of the proposed method.
    
\item {\bf Code of ethics}
    \item[] Question: Does the research conducted in the paper conform, in every respect, with the NeurIPS Code of Ethics \url{https://neurips.cc/public/EthicsGuidelines}?
    \item[] Answer: \answerYes{}
    \item[] Justification: The research complies with the NeurIPS Code of Ethics. 

\item {\bf Broader impacts}
    \item[] Question: Does the paper discuss both potential positive societal impacts and negative societal impacts of the work performed?
    \item[] Answer: \answerYes{}
    \item[] Justification: We discuss both positive and negative societal impacts in Appendix~\ref{sec:app_A3}. 
    
\item {\bf Safeguards} 
    \item[] Question: Does the paper describe safeguards that have been put in place for responsible release of data or models that have a high risk for misuse (e.g., pre-trained language models, image generators, or scraped datasets)?
    \item[] Answer: \answerNA{}{}
    \item[] Justification: The proposed method is a defensive perturbation technique and does not constitute a high-risk model or dataset for misuse as defined in the guidelines.

\item {\bf Licenses for existing assets}
    \item[] Question: Are the creators or original owners of assets (e.g., code, data, models), used in the paper, properly credited and are the license and terms of use explicitly mentioned and properly respected?
    \item[] Answer: \answerYes{}
    \item[] Justification: We properly cite all datasets, models, and prior works used in this paper.

\item {\bf New assets}
    \item[] Question: Are new assets introduced in the paper well documented and is the documentation provided alongside the assets?
    \item[] Answer: \answerNA{}.
    \item[] Justification: This paper does not release new assets.

\item {\bf Crowdsourcing and research with human subjects}
    \item[] Question: For crowdsourcing experiments and research with human subjects, does the paper include the full text of instructions given to participants and screenshots, if applicable, as well as details about compensation (if any)? 
    \item[] Answer: \answerYes{}
    \item[] Justification: We provide the full instruction text, screenshots of the experimental interface, and details about participant compensation in Appendix~\ref{sec:app_D}.

\item {\bf Institutional review board (IRB) approvals or equivalent for research with human subjects}
    \item[] Question: Does the paper describe potential risks incurred by study participants, whether such risks were disclosed to the subjects, and whether Institutional Review Board (IRB) approvals (or an equivalent approval/review based on the requirements of your country or institution) were obtained?
    \item[] Answer: \answerYes{}
    \item[] Justification: The user study involved minimal risk, as participants were only asked to watch and evaluate short video clips without exposure to sensitive or harmful content. Under our institutional guidelines, such minimal-risk studies do not require formal IRB approval.

\item {\bf Declaration of LLM usage}
    \item[] Question: Does the paper describe the usage of LLMs if it is an important, original, or non-standard component of the core methods in this research? Note that if the LLM is used only for writing, editing, or formatting purposes and does \emph{not} impact the core methodology, scientific rigor, or originality of the research, declaration is not required.
    \item[] Answer: \answerNA{}
    \item[] Justification: The core method does not involve the use of LLMs.

\end{enumerate}

\end{document}